
\font\BBBig=cmr10 scaled\magstep3
\magnification 1200


\def\foot#1{
\footnote{($^{\the\foo}$)}{#1}\advance\foo by 1
} 


\def\title{
{\bf\BBBig
\centerline{On Schr\"odinger superalgebras
}\bigskip
}}

\def\runningtitle{
Super-Schr\"odinger
} 


\def\authors{
\centerline{C.~DUVAL\foot{D\'epartement de Physique,
UFR de Luminy, Universit\'e d'Aix-Marseille II and
Centre de Physique Th\'eorique
CNRS, Case 907, F--13288 MARSEILLE
Cedex 09 (France). 
}
and 
P. A. HORV\'ATHY\foot{D\'epartement de Math\'ematiques,
Universit\'e de Tours, Parc de Grandmont,
F--37200 TOURS (France), 
}}
}

\def\runningauthors{
Christian DUVAL and P\'eter HORV\'ATHY
} 


\voffset = 1cm 
\baselineskip = 16pt 

\headline ={
\ifnum\pageno=1\hfill
\else\ifodd\pageno\hfil\tenit\runningtitle\hfil\tenrm\folio
\else\tenrm\folio\hfil\tenit\runningauthors\hfil
\fi\fi
} 

\nopagenumbers
\footline = {\hfil} 



\font\tenb=cmmib10 
\newfam\bsfam
 
\textfont\bsfam=\tenb

\mathchardef\betab="080C
\mathchardef\xib="0818
\mathchardef\omegab="0821
\mathchardef\deltab="080E
\mathchardef\epsilonb="080F
\mathchardef\pib="0819
\mathchardef\sigmab="081B
\mathchardef\zetab="0810

\mathchardef\balpha="080B
\mathchardef\bbeta="080C
\mathchardef\bgamma="080D
\mathchardef\bomega="0821

\def\={{\!=}}
\def\parag{\hfill\break} 
\def\and{\qquad\hbox{and}\qquad}
\def\where{\qquad\hbox{where}\qquad}
\def\kikezd{\parag\underbar}
\def\cJ{{\cal J}}
\def\cB{{\cal G}}
\def\cG{{\cal G}}
\def\cP{{\cal P}}
\def\cM{{\cal M}}
\def\cD{{\cal D}}
\def\cK{{\cal K}}
\def\cL{{\cal L}}
\def\cH{{\cal H}}
\def\cQ{{\cal Q}}
\def\cS{{\cal S}}
\def\cY{{\cal Y}}
\def\cZ{{\cal Z}}

\def\bP{{\bf P}} 
\def\bQ{{\bf Q}} 
\def\bR{{\bf R}}

\def\bV{{\bf V}}
\def\bW{{\bf W}}
\def\br{{\bf r}}
\def\bp{{\bf p}}
\def\bq{{\bf q}}

\def\bPsi{{\bf\Psi}}
\def\bXi{{\bf\Xi}}
\def\smallcirc{{\raise 0.5pt \hbox{$\scriptstyle\circ$}}}
\def\smallover#1/#2{\hbox{$\textstyle{#1\over#2}$}}
\def\half{{\smallover 1/2}}
\def\2{{\smallover 1/2}}
\def\between#1{\qquad\hbox{#1}\qquad}

\def\semidirectproduct{
{\ooalign
{\hfil\raise.07ex\hbox{s}\hfil\crcr\mathhexbox20D}}
}

\def\ccr{\cr\noalign{\medskip}}

\def\N{N_+,N_-}
\def\m{m_+,m_-}


\newcount\ch 
\newcount\eq 
\newcount\foo 
\newcount\ref 

\def\chapter#1{
\bigskip\goodbreak\parag\eq = 1\advance\ch by 1{\bf\the\ch.\enskip#1}
}

\def\equation{
\leqno(\the\ch.\the\eq)\global\advance\eq by 1
}

\def\reference{
\parag ${}^{\number\ref}$ \advance\ref by 1
}

\ch = 0 
\foo = 1 
\ref = 1 


\def\math#1{\mathop{\rm #1}\nolimits}

\def\const{\math{const}}

\def\sp{\math{sp}}
\def\osp{\math{osp}}
\def\o{\rm o}
\def\sch{\math{sch}}
\def\Tr{\math{Tr}}


\title
\vskip 1.0cm
\authors
\vskip .25in
\vskip 0.5in

\centerline{Abstract.}

\noindent
{\it
We construct, using the supersymplectic framework of Berezin, Kostant and
others, two types of supersymmetric extensions of the Schr\"odinger algebra
(itself a conformal extension of the Galilei algebra).
An `$I$-type' extension exists in any space dimension, and for any pair of
integers $N_+$ and $N_-$. It yields an $N=N_++N_-$ superalgebra, which
generalizes the $N=1$ supersymmetry Gauntlett et al. found for a free
spin-$\half$ particle, as well as the $N=2$ supersymmetry of the fermionic
oscillator found by Beckers et al. In two space dimensions, new, `exotic' or
`$IJ$-type' extensions arise for each pair of integers $\nu_+$ and $\nu_-$,
yielding an $N=2(\nu_++\nu_-)$ superalgebra of the type discovered recently by
Leblanc et al. in non relativistic Chern-Simons theory.
For the magnetic monopole the symmetry reduces to
$\o(3)\times\osp(1/1)$,
and for the magnetic vortex it reduces to
$\o(2)\times\osp(1/2)$.
}

\vskip1cm
October 1993

CPT--93/P.2912 
{\sl Journ. Math. Phys}. {\bf 35} 2516 (1994).



\chapter{Introduction.}

Recent interest in non-relativistic supersymmetries stems
from Chern-Simons theory: as found by Leblanc, Lozano and Min (LLM)~[1], 
non-relativistic Chern-Simons theory in two dimensions
admits an $N=2$ conformal supersymmetry, which extends the Schr\"odinger
symmetry discovered previously by Jackiw et al.~[2].

The story starts in the early seventies, when Niederer, and Hagen,
[3] pointed out that the maximal kinematical invariance group of the
free, spin-$0$, Schr\"odinger equation forms a $13$-dimensional Lie group
whose Lie algebra is now
called the (extended) {\it Schr\"odinger algebra} $\widetilde{\sch}(3)$. 
This latter contains, in addition to the (centrally extended) Galilei group, 
two `conformal' generators, namely dilatations, $\cD$, 
and expansions, $\cK$, which close with the Hamiltonian, $\cH$, into an
$\math{sl}(2,\bR)\cong\o(2,1)$ subalgebra~[4].
 
Gauntlett et al.~[5] extended this algebra to 
spin-$\half$ particles with spin generator $\zetab$. 
They found five conserved odd generators, namely
$$
\left\{
\matrix{
\cQ
&=&
{\displaystyle{1\over\sqrt{m}}\,\bp\cdot\zetab}\hfill
&\hbox{(helicity)}\hfill
\cr
\cS
&=&
\sqrt{m}\,\left({\displaystyle\br-{\bp\over m}t}\right)\cdot\zetab
\qquad\hfill
&\hbox{(super-expansion)}\hfill 
\ccr
{\bXi}&=&\sqrt{m}\,\zetab
\hfill
&\hbox{(spin)}\hfill 
\cr
}\right.
\equation
$$
(where $\bp=\dot\br$), which provide an $18$-dimensional, 
$N=1$ supersymmetric extension of the Schr\"odinger algebra.

Soon after the discovery of the Schr\"odinger symmetry of the free particle,
Niederer~[6] found that the harmonic oscillator (described by the
bosonic Hamiltonian $H_B$ below) also admits this same Schr\"odinger 
symmetry ---~although the generators look quite different, cf. Eq.~(7.5). 
This result has been extended by Beckers et al.~[7] to the $n$-dimensional
fermionic oscillator with total Hamiltonian
$$
H_{\rm tot}=H_B+H_F=\half\left({\bp^2\over m}+m\omega^2\br^2\right)
-i\omega\sum_{a=1}^n\left(\zeta_+^a\zeta_-^a-\zeta_-^a\zeta_+^a\right),
\equation
$$
the $\zeta^a_\pm$ ($a=1,\ldots,n)$ being the generators of a Clifford algebra. 
This system has an $N=2$ conformal supersymmetry, with supercharges
$$
\left\{
\eqalign{
&Q_\pm= \left(\bp\mp im\omega \br\right)\cdot{\zetab_\pm\over\sqrt{m}}
\ccr
&S_\pm=e^{\mp2i\omega t}
\left(\bp\pm im\omega \br\right)\cdot{\zetab_\pm\over\sqrt{m}}
\ccr
&{\bf T}_\pm=
e^{\mp i\omega t}{\sqrt{m}}\ \zetab_\pm. 
\cr
}\right.
\equation
$$
Note that $H_B$ and $H_F$ are both bosonic and are separately conserved. 
For $n=3$, this algebra has $21$ generators.

\goodbreak

Non-relativistic Chern-Simons theory in $2+1$ dimensions also
has a Schr\"odinger symmetry~[2], extended by Leblanc et al.~[1] into a
$16$-dimensional $N=2$ supersymmetry algebra.
But, in two space dimensions, the `ordinary' $N=2$ superconformal symmetry of
the type of Beckers et al.~[7] has $18$ generators: in particular, there are
$N=2$ of each of the `$\cQ$-type', `$\cS$-type' and `${\bf T}$-type' charges.
In the LLM algebra, however,  there are $N=2$ `$\cQ$-type' and `$\cS$-type'
supercharges, but just  $N/2=1$ `${\bf T}$-type' charge.
The commutation relations ot the LLM-algebra are similar to, but still
different from, those of~[7] for $n=2$.

 The main result of this paper is to confirm the `exotic' supersymmetry of
Leblanc et al.: we construct, within the symplectic framework~[8], extended to
the Grassmann case  by Berezin, Katz, Kostant, and others~[9,10], {\it two
types} of supersymmetric extensions of the Schr\"odinger algebra. Our clue is
to first imbed the Schr\"odinger algebra $\widetilde{\sch}(n)$ into
$\widetilde{\math{isp}}(n)$, the (central extension of the) algebra of
inhomogeneous (or affine) symplectic transformations. But this latter admits a
natural supersymmetric extension, namely the affine orthosymplectic algebra
$\widetilde{\math{iosp}}(n/m)$ obtained by adding $m$ odd variables.
We look therefore for supersymmetric extensions of $\sch(n)$ {\it within} 
$\widetilde{\math{iosp}}(n/m)$. This is reviewed in the first part of the
article.

Our $I${\it -type} extension $\widetilde{\sch}(n/N_+,N_-)$
exists for any spatial dimension $n$.
It is labelled by a pair integers $N_+$ and $N_-$, related to the dimension
and the signature of the Grassmann space.
Our algebra has $N=N_++N_-$ of each of the `Q-type', 
`S-type' and `${\bf T}$-type' charges.
If $N_-=0$ ($N=N_+$), we denote this algebra simply by 
$\widetilde{\sch}(n/N)$.

If, however, space is {\it two-dimensional}, we have another `{\it exotic}'
or `$IJ$-{\it type}' extension, we denote by
$\widetilde{\sch}_e(\nu_+,\nu_-)$.
It is again labelled by two integers, $\nu_+$ and $\nu_-$, and yields
$N=2\nu=2(\nu_++\nu_-)$ of each of the `Q-type and `S-type' charges but only
$\nu$ `${\bf T}$-type' charges. 

In the second part of our paper, we illustrate our theory on examples.

-- Firstly, we recover the $N=1$ super-Schr\"odinger algebra of Gauntlett et
al.~[5], in the $I$-type framework.

-- Beckers et al.'s oscillator superalgebra~[7] corresponds to an
$I$-type extension with $N=N_+=2$.

-- The superalgebra of Leblanc et al.~[1] is obtained as the
`exotic' (`$IJ$-type') extension $\widetilde{\sch}_e(\nu_+=1,\nu_-=0)$.

-- Parts of the symmetry algebra may remain unbroken by certain
interactions. A celebrated example is the $\o(2,1)$ symmetry of the
Dirac monopole~[11], which extends to $\osp(1/1)$ when spin is added~[12].
Another, more recent, example is provided by the magnetic vortex, which also
carries the $\o(2,1)$ bosonic symmetry~[13]. Now, the Hamiltonian admits
{\it two} distinct square roots~[14] that combine with the bosonic algebra
into an $\osp(1/2)$ sub-superalgebra of our `exotic' super-Schr\"odinger
algebra~[15] as illustrated by section~9.

\goodbreak


\chapter{The orthosymplectic algebra.}

Let us start with a supervector space $E=E_0\oplus E_1$; the 
endomorphisms of $E$ admit the even/odd decomposition
$$
Z=\pmatrix{
A&B\cr
C&D\cr
}
=
\pmatrix{
A&0\cr
0&D\cr
}
+
\pmatrix{
0&B\cr
C&0\cr
}.
\equation
$$
The {\it supercommutator} which is defined on homogeneous endomorphisms by
$[Z,Z'] = ZZ'-(-1)^{|Z|\cdot|Z'|}Z'Z$ reads
$$\eqalign{
\left[
\pmatrix{
A&B\ccr
C&D\cr
}\right.,&\left.
\pmatrix{
A'&B'\ccr
C'&D'\cr
}
\right]
=\ccr
&\pmatrix{
AA'-A'A+BC'+B'C
&
BD'-B'D+AB'-A'B
\ccr
CA'-C'A+DC'-D'C
&
DD'-D'D+CB'+C'B\cr
}.
\cr
}
\equation
$$
From now on we will be dealing with $E=\bR^{2n}\oplus\bR^m_1\equiv\bR^{2n/m}$,
globally parametrized by
$$
X=\pmatrix{\bP\cr\bQ\cr}\in\bR^{2n}
\and
\xi = \pmatrix{\xi^1\cr\vdots\cr\xi^m\cr}\in\bR^m_1
$$
where the $\xi^\alpha$
($\alpha=1,\ldots,m)$ are, along with $1$, the generators of the Grassmann
algebra $\Lambda\bR^m$ and $\bR^m_1\equiv\Lambda^1\bR^m$.
Now $\bR^{2n/m}$ carries the canonical symplectic structure of prescribed
signature $s$ given by~[9,10]
$$
\omega =
\sum_{j=1}^n{dP^j\wedge dQ^j}+
\half\sum_{\alpha=1}^m{\varepsilon_\alpha\,d\xi^\alpha\wedge d\xi^\alpha},
\equation
$$
where $\varepsilon_\alpha=\pm 1$ and
$s=\sum{\varepsilon_\alpha}$.
Using the anticommutativity of the odd variables, it is easy to check that

a)
$\omega$ is superskewsymmetric,
i.e. $\omega(V,V')+(-1)^{|V|\cdot|V'|}\,\omega(V',V)=0$
for all homo\-geneous elements $V,V'\in\bR^{2n/m}$;

b)
$\omega$ is nondegenerate since it is represented by the even
invertible endomorphism
$$
\pmatrix{
J&0\cr
0&G\cr
}
$$
according to
$\omega(V_0+V_1,V'_0+V'_1)=V^T_0JV'_0+V^T_1GV'_1$
where
$$
J=\pmatrix{
0&I_n\ccr
-I_n&0\cr
}
\between{and}
G=\pmatrix{
\varepsilon_1 &&\cr
\noalign{\vskip -1mm}
&\ddots&\cr
&&\varepsilon_m\cr
};
$$

c)
$\omega$ is exact:
$$
\omega=\half\,d\left(
\overline{\bP}\,d\bQ - d\overline{\bP}\,\bQ +
\overline{\xi}\,d\xi\right).
\equation
$$
In the last expression, the `overline' stands for the adjoint with respect
to a metric, e.g. $\overline{\bP}\equiv\bP^T$ and 
$\overline\xi\equiv\xi^T G$.
We will systematically use this notation in the sequel.\foot{If $M\in
L((\bR^p,G)$, $(\bR^q,H))$ then $\overline{M}\equiv G^{-1}M^T H$.}

Let us now recall the definition~[9-11] of the {\it orthosymplectic algebra}
$\osp(n/m)$ which consists of those endomorphisms
$Z$ of $\bR^{2n/m}$ whose homogeneous components verify
$
\omega(ZV,V')+(-1)^{|Z|\cdot|V|}\,\omega(V,ZV')=0
$
for all homogeneous elements $V,V'\in\bR^{2n/m}$.
In other words, an endomorphism $Z$ of the form $(2.1)$ is orthosymplectic
for the super\-symplectic structure
$
\omega
=
\half\left(d\overline{X}\wedge JdX + d\overline{\xi}\wedge d\xi\right)
$
of $\bR^{2n/m}$ with signature $s=m_+-m_-$ if
$$
d(\delta\overline{X})\wedge JdX +
d\overline{X}\wedge Jd(\delta{X})+d(\delta\overline\xi)
\wedge d\xi+
d\overline\xi\wedge d(\delta\xi)
=0
$$
where
$\delta{X}=AX+B\xi,
\delta\xi=CX+D\xi$,
that is if $\overline{A}J+JA=0$, $C=\overline{B}J$ and
$\overline{D}+D=0$. We get therefore the superalgebra
$$
\osp(n/\m)=
\left\{
\pmatrix{
A&B\ccr
\overline{B}J&D\ccr
}
\Bigm| A\in\sp(n,\bR), B\in L(\bR^m_1,\bR^n), D\in\o(\m)
\right\}
\equation
$$
whose multiplication law is given by (2.2).



\chapter{The Schr\"odinger algebra.}

We first recover the Schr\"odinger algebra~[3,4] in a novel way,
namely as a subalgebra of the extended affine symplectic algebra
$\widetilde{\math{isp}}(n)$. This is achieved by identifying $\bR^{2n}$ with 
$\bR^{n}\otimes\bR^{2}$ (i.e. with the $n\times 2$ matrices) via  
$$
\pmatrix{
\bP\cr
\bQ\cr
}
\longmapsto (\bP\bQ).
$$

One shows now that the linear transformations 
$$
(\bP\bQ)\longmapsto A(\bP\bQ)B^{-1}
\equation
$$ 
are
symplectic if $A\in\o(n)$ and $B\in\math{SL}(2,\bR)$.
The corresponding Lie algebra is isomorphic to
$$
\left\{
\pmatrix{
A+aI_n&bI_n\ccr
cI_n&A-aI_n\cr
}
\Bigm|
A\in\o(n);a,b,c\in\bR
\right\}
\cong
\o(n)\times\math{sl}(2,\bR)
\subset
\sp(n,\bR).
\equation
$$

Note that the `dual pair' $\o(n)\times\math{sl}(2,\bR)$
is the {\it homogeneous} Schr\"odinger algebra~[3,4]. 
The full centrally extended Schr\"odinger Lie algebra~[3,4] is
conveniently obtained~[8,12] by considering rather the
canonical contact structure on $\bR^{2n+1}$, 
$$
\varpi=\half\left(\overline\bP\,d\bQ-d\overline\bP\,\bQ\right)+ds,
\equation
$$
and look for the infinitesimal affine contact transformations 
$$
\pmatrix{\bP\cr\bQ\cr s\cr}\longmapsto\delta\pmatrix{\bP\cr\bQ\cr s\cr}
$$
which Lie-transport the $1$-form $\varpi=0$, 
$$
\half\left(
\delta\overline\bP)\,d\bQ
+\overline\bP\,d(\delta\bQ)
-d\overline{\bP}\,(\delta\bQ)
-d(\delta\overline\bP)\,\bQ
\right)
+d(\delta{s})
=0,
$$
and {\it extend} $\o(n)\times\math{sl}(2,\bR)$. One readily
finds
$$
\delta\pmatrix{
\bP\ccr
\bQ\ccr
s\ccr
1\cr
}
=
\pmatrix{
A+aI_n & bI_n & 0 & \bV \ccr
cI_n & A-aI_n & 0 & \bW \ccr
\half\overline\bW & -\half\overline\bV & 0 & u\ccr
0 & 0 & 0 & 0\cr
}
\pmatrix{
\bP\ccr
\bQ\ccr
s\ccr
1\cr
}
\equation
$$
where $A\in\o(n);a,b,c,u\in\bR;\bV,\bW\in\bR^n$
which defines the extended Schr\"odinger algebra $\widetilde{\sch}(n)$,
the central $\bR$-extension of
$$
\sch(n)\cong\left(\o(n)\times\math{sl}(2,\bR)\right)
\,\semidirectproduct\,\bR^{2n}.
\equation
$$



\chapter{Schr\"odinger superalgebras: $I$-type extensions.}

Let us now merge the two previous structures into a single one, {\it viz.} 
the canonical contact structure on $\bR^{2n+1/m}$~[10] given by the even
$1$-form
$$
\alpha
=
\half\left(\overline\bP\,d\bQ -d\overline\bP\,\bQ\right)
+
\half\,\overline\xi\,d\xi+ds
\equation
$$
such that $\omega=d\alpha$.
A tedious but routine calculation shows that the infinitesimal
affine contact transformations are given by
$$
\normalbaselineskip=15pt
\delta\pmatrix{
\bP\cr
\bQ\cr
s\cr
1\cr
\xi\cr
}
=
\pmatrix{
A& B& 0& \bV &-\Sigma\cr
C& -\overline{A}&0&\bW &-\Theta\cr
\half\overline\bW&-\half\overline\bV&0& u&-\half\overline\Psi\cr
0 & 0 & 0 & 0 & 0\cr
\overline\Theta&-\overline\Sigma&0&\Psi&D\cr
}
\pmatrix{
\bP\cr
\bQ\cr
s\cr
1\cr
\xi\cr
}
\equation
$$
with $B=\overline{B}$, $C=\overline{C}$,
$u\in\bR$; $\bV,\bW\in\bR^n$; 
$\Sigma,\Theta\in{L(\bR^m_1,\bR^n)}$,
$\Psi\in\bR^m_1$ 
and $\overline{D}+D=0$,
i.e.
$D\in\o(\m)$. This is, indeed, the {\it extended affine
orthosymplectic
superalgebra} 
$\widetilde{\math{iosp}}(n/m_+,m_-)$,
spanned by the vector spaces 
$\widetilde{\math{isp}}(n,\bR)$,
$\o(m_+,m_-)$,
and
$L(\bR_1^m,\bR^{2n+1})$.

Let us now look for Lie superalgebras that 
extend nontrivially the Schr\"odinger algebra 
$\widetilde{\math{sch}}(n)$ in (3.4) {\it within} 
$\widetilde{\math{iosp}}(n/m_+,m_-)$. Thus, the commutator of two
infinitesimal transformations (4.2) should close according to (3.4);
the crucial condition is that $A$, $B$ and $C$ in
Eq.~(4.2) be of the same form as in Eq.~(3.4).
By Eq.~(2.2), this requires 
$$
\left\{
\eqalign{
&\Sigma\overline{\Theta'}+\Sigma'\overline{\Theta}+
\Theta\overline{\Sigma'}+\Theta'\overline{\Sigma}=a''I_n
\ccr
&\Sigma\overline{\Sigma'}+\Sigma'\overline{\Sigma}=b''I_n
\ccr
&
\Theta\overline{\Theta'}+\Theta'\overline{\Theta}=c''I_n
\cr}\right.
\equation
$$
for some scalars $a''$, $b''$ and $c''$. 
One way of satisfying these conditions is to assume that
the matrices split into $N$ blocks,
$$
G=\pmatrix{
\epsilon_1I_n&&\cr
&\ddots&\cr
&&\epsilon_NI_n\cr
},
\equation
$$
so that $m=nN$ and 
$
\xi=\pmatrix{\xib_1\cr\vdots\cr\xib_N\cr}
$
with 
$\xib_j\in\bR_1^n$,
supplemented by the `$I$-type' Ansatz
$$
\Sigma=(\sigma^1I_n\;\ldots\;\sigma^NI_n)
\and
\Theta=(\theta^1I_n\;\ldots\;\theta^NI_n),
\equation
$$
where $\sigma^j,\theta^k\in\bR_1$ for all $j,k=1,\ldots,N$.

\goodbreak

Let $N_+$ (resp. $N_-$) be the number of blocks in (4.4) with positive (resp.
negative) sign and $N=N_+ + N_-$. For each pair $(N_+,N_-)$, we get a
supersymmetric extension of the Schr\"odinger algebra we call 
$(N_+,N_-)$-{\it Schr\"odinger superalgebra of $I$-type}, and denote by 
$\widetilde{\sch}(n/N_+,N_-)$. It consists of endomorphisms in
$\math{sl}(2n+2/nN)$ of the form
$$
Z=
\pmatrix{
A+aI_n&bI_n&0&\bV&-\sigma^1I_n&\ldotp&\ldots&-\sigma^NI_n
\ccr
cI_n &A-aI_n&0&\bW&-\theta^1I_n&\ldotp&\ldots&-\theta^NI_n
\ccr
\half\overline\bW&-\half\overline\bV&0&u&
-\half\overline{\bPsi}{}^1&\ldotp&\ldots&-\half\overline{\bPsi}{}^N
\ccr
0&0&0&0&0&\ldotp&\ldots&0
\ccr
\theta_1I_n&-\sigma_1I_n&0&\bPsi_1&A&R_1^2I_n&\ldots&R_1^NI_n
\ccr
\ldotp&\ldotp&\ldotp&\ldotp&R_2^1I_n&A&\ldots&\ldotp
\ccr
\vdots&\vdots&\vdots&\vdots&\vdots&\vdots&\ddots&\vdots
\ccr
\theta_NI_n&-\sigma_NI_n&0&\bPsi_N&
R_N^1I_n&\ldotp&\ldots&A
\ccr
}
\equation
$$
with
$A\in\o(n)$;
$a,b,c,u\in\bR$;
$\bV,\bW\in\bR^n$;
$R=(R_j^k)\in\o(\N)$;
$\sigma_j\equiv\epsilon_j\sigma^j,
\theta_j\equiv\epsilon_j\theta^j\in\bR_1$,
$\bPsi_j\in\bR^n_1$,
$\overline{\bPsi}{}^j\equiv\epsilon_j\bPsi_j^T$
(all $j,k=1,\ldots,N$).
The supercommutators $Z''=[Z,Z']$ read
\goodbreak
$$
\left\{\eqalign{
A''&=AA'-A'A
\cr
a''&=bc'-b'c-
\sum_{j=1}^N{\left(\theta_j{\sigma'}^j+\theta'_j\sigma^j\right)}
\cr
b''&=2(ab'-a'b)+2\sum_{j=1}^N{\sigma_j{\sigma'}^j}
\cr
c''&=2(ca'-c'a)-2\sum_{j=1}^N{\theta_j{\theta'}^j}
\cr
\bV''&=A\bV'-A'\bV+a\bV'-a'\bV+b\bW'-b'\bW 
-\sum_{j=1}^N{\left(\bPsi_j{\sigma'}^j+\bPsi'_j\sigma^j\right)}
\cr
\bW''&=A\bW'-A'\bW-a\bW'+a'\bW+c\bV'-c'\bV 
-\sum_{j=1}^N{\left(\bPsi_j{\theta'}^j+\bPsi'_j\theta^j\right)}
\cr
u''&=-\overline\bV\,\bW'+\overline{\bV'}\,\bW 
-\sum_{j=1}^N{\overline{\bPsi}{}^j\bPsi'_j}
\cr
{R''}_j^k&=-\sum_{i=1}^N{\left(R_i^k{R'}_j^i-{R'}_i^kR_j^i\right)}
+\theta^k\sigma_j'-{\theta'}^k\sigma_j
-\sigma^k\theta_j'-{\sigma'}^k\theta_j
\cr
}\right.
\equation
$$
for the bosonic part, and
$$
\left\{
\eqalign{
{\sigma''}^j
&=
a\sigma^j-a'\sigma^j+b{\theta'}^j-b'\theta^j
-\sum_{k=1}^N{\left(R_k^j{\sigma'}^k-{R'}_k^j\sigma^k\right)},
\cr
{\theta''}^j
&=
-a{\theta'}^j+a'\theta^j+c{\sigma'}^j-c'\sigma^j
-\sum_{k=1}^N{\left(R_k^j{\theta'}^k-{R'}_k^j\theta^k\right)},
\ccr
\bPsi_j''
&=
A\bPsi_j'-A'\bPsi_j-\sigma_j\bW'+\sigma_j'\bW+\theta_j\bV'-\theta_j'\bV
-\sum_{k=1}^N{\left(\bPsi_k{R'}_j^k-\bPsi_k'R_j^k\right)}.
\cr}\right.
\equation
$$
for the fermionic part.

The components of the associated supermoment map~[10],
$$
\mu:\bR^{2n+1/nN}\longrightarrow\widetilde{\sch}(n/N_+,N_-)^*,
$$ 
are found using the expression
$$
\langle\mu,Z\rangle=\imath(Z_E)\alpha,
$$
the infinitesimal action $Z_E$ of $Z$
(given by (4.6)) on $E=\bR^{2n+1/nN}$ being defined in (4.2).
Putting
$$
\eqalign{
\langle \mu,Z\rangle=&\half\Tr(\cJ A)+\cH c
-\cD a-\cK b
-\overline\cB\,\bV
+\overline\cP\,\bW
+\cM u
-\half\,\cH_{jk}R^{jk}
\ccr
&
+\cQ_j\theta^j 
-\cS_j\sigma^j
+\overline{\bXi}{}^j\bPsi_j
\cr
}
\equation
$$
where we have adopted the Einstein summation convention and used the metric
$$
g_{jk}=\epsilon_j\delta_{jk}
$$
to raise and lower fermionic indices, we
claim that
$$
\left\{
\matrix{
\cJ\hfill 
&=&\bQ\,\overline\bP-\bP\,\overline{\bQ}
-\xib_j\,\overline{\xib}{}^j,\qquad
\qquad\hfill
&\cQ_j\hfill &=&\overline{\bP}\,\xib_j,\hfill
\ccr
\cH\hfill &=&\half ||\bP||^2,\hfill
&\cS_j\hfill &=&\overline\bQ\,\xib_j,\hfill
\ccr
\cD\hfill &=&\overline\bP\,\bQ,\hfill
&\bXi_j\hfill &=&\xib_j,\hfill
\ccr
\cK\hfill &=&\half ||\bQ||^2,\hfill&
\ccr
\cB\hfill &=&\bQ,\hfill&
\ccr
\cP\hfill &=&\bP,\hfill& 
\ccr
\cM\hfill &=&1,\hfill
&\ccr
\cH^{j}_k&=&\overline{\xib}{}^j\,\xib_k,\hfill&
\cr
}\right.
\equation
$$
where $j,k=1,\ldots,N$.
The components of $\mu$ plainly generate, under {\it super-Poisson brackets}
we still denote by $[\,\cdot\,,\,\cdot\,]$,
a superalgebra isomorphic to $\widetilde{\sch}(n/\N)$.

\goodbreak

The crucial formula\foot{According to our convention,
$\left[Z_E,Z'_E\right]\equiv-[Z,Z']_E$.}
$$
\big[\langle \mu,Z\rangle,\langle \mu,Z'\rangle\big]
=
-\langle \mu,[Z,Z']\rangle
\equation
$$
yields, using the commutation relations (4.7,8), 
the `even' Poisson brackets
$$
[\cJ_{ab},\cJ_{cd}]\hfill=
\delta_{cb}\cJ_{ad}-\delta_{ca}\cJ_{bd}+
\delta_{db}\cJ_{ca}-\delta_{da}\cJ_{cb},
\equation
$$
and
$$\matrix{
[\cH,\cD]\hfill&=&2\cH\;\hfill
&[\cH,\cK]\hfill&=&\cD\hfill
&[\cD,\cK]\hfill&=&2\cK\hfill&
\ccr
[\cJ_{ab},\cH]\hfill&=&0\hfill
&[\cJ_{ab},\cD]\hfill&=&0\hfill
&[\cJ_{ab},\cK]\hfill&=&0\hfill
\ccr
[\cB_a,\cB_b]\hfill&=&0\hfill
&[\cP_a,\cP_b]\hfill&=&0\hfill
&[\cP_a,\cB^b]\hfill&=&\delta_a^b\cM\hfill
\ccr
[\cB_a,\cJ^{bc}]\hfill&=&\delta_a^b\cB^c-\delta_a^c\cB^b\hfill
&[\cP_a,\cJ^{bc}]\hfill&=&\delta_a^b\cP^c-\delta_a^c\cP^b\hfill&
\ccr
[\cH,\cB_a]\hfill&=&\cP_a\hfill
&[\cH,\cP_a]\hfill&=&0\hfill
&[\cD,\cB_a]\hfill&=&\cB_a\hfill
\ccr
[\cD,\cP_a]\hfill&=&-\cP_a\hfill
&[\cK,\cB_a]\hfill&=&0\hfill
&[\cK,\cP_a]\hfill&=&-\cB_a,\hfill
\cr
}
\equation
$$
where $a, b, c, d=1,\ldots,n$, ($\cM$ commutes with everything).
These are the commutation relations of
$\left(\o(n)\times\o(2,1)\right)
\,\semidirectproduct\,
\rm{h}(n)$,
i.e. that of the extended Schr\"odinger algebra $\widetilde{\sch}(n)$.
The $\cH_{ij}$'s 
commute with all other bosonic generators and, themselves, 
generate $\o(\N)$:
$$
[\cH_{ij},\cH_{k\ell}]\hfill=
-g_{kj}\cH_{i\ell}+g_{ki}\cH_{j\ell}
-g_{\ell j}\cH_{ki}+g_{\ell i}\cH_{kj}
\equation
$$
where $i,j,k,\ell=1,\ldots,N$.
The bosonic commutation relations are hence those of the direct product
of the extended Schr\"odinger algebra with $\o(N_+,N_-)$. They are
supplemented with the relations involving `fermionic' (i.e. odd)
component. Firstly, we have
$$
\matrix{
[\cQ_j,\cD]\hfill&=&\cQ_j\hfill
&[\cQ_j,\cK]\hfill&=&\cS_j\hfill
\ccr
[\cQ_j,\cH]\hfill&=&0\hfill
&[\cQ_i,\cH_{jk}]\hfill&=&g_{ik}\cQ_j-g_{ij}\cQ_k\hfill
\ccr
[\cS_j,\cD]\hfill&=&-\cS_j\hfill
&[\cS_j,\cK]\hfill&=&0\hfill
\ccr
[\cS_j,\cH]\hfill&=&-\cQ_j\hfill
&[\cS_i,\cH_{jk}]\hfill&=&g_{ik}S_j-g_{ij}\cS_k\hfill
\ccr
[\cQ_j,\cQ_k]\hfill&=&-2g_{jk}\cH\hfill
&[\cS_j,\cS_k]\hfill&=&-2g_{jk}\cK\hfill
\ccr
[\cQ_j,\cS_k]\hfill&=&-g_{jk}\cD+\cH_{jk},\qquad\hfill
&&&\hfill
\ccr}
\equation
$$
which says that the $\cH,\cD,\cK,\cH_{jk}$ close with $\cQ_j,\cS_j$
into $\osp(1/\N)$.

Next, according to
$$
[\cQ_j,\cJ_{ab}]=0,
\qquad
\qquad
[\cS_j,\cJ_{ab}]=0,
\equation
$$
the `homogeneous super-Schr\"odinger algebra' is the direct 
product $\o(n)\times\osp(1/\N)$.
The commutators of translations and super-translations,
$$
[\cP_a,\Xi_j^b]=0,
\qquad
[\cB_a,\Xi_j^b]=0,
\qquad
[\Xi_j^a,\Xi_k^b]=
-g_{jk}\,\delta^{ab}\cM,
\equation
$$
define the super-Heisenberg algebra ${\rm h}(n/\N)$ Beckers et al. denote 
by ${\rm sh}(n)$.
Finally,
$$
\matrix{
[\Xi_j^a,\cJ^{bc}]\hfill&=&
\delta^{ab}\,\Xi_j^c-\delta^{ac}\,\Xi_j^b\qquad\hfill
&&&
\ccr
[\cH,\Xi_j^a]\hfill&=&0\hfill
&[\cD,\Xi_j^a]\hfill&=&0\hfill
\ccr
[\cK,\Xi_j^a]\hfill&=&0\hfill
&[\cH_{jk},\Xi_i^a]\hfill&=&
g_{ij}\,\Xi_k^a-g_{ik}\,\Xi_j^a\quad\hfill
\ccr
[\cQ_j,\cB^a]\hfill&=&\Xi_j^a\hfill
&[\cQ_j,\cP]\hfill&=&0\hfill
\ccr
[\cS_j,\cB_a]\hfill&=&0\hfill
&[\cS_j,\cP^a]\hfill&=&-\Xi_j^a\hfill
\ccr
[\cQ_j,\Xi_k^a]\hfill&=&-g_{jk}\,\cP^a\hfill
&[S_j,\Xi_k^a]\hfill&=&-g_{jk}\,\cB^a\hfill
\ccr
}
\equation
$$
fixes the relation between the homogeneous and inhomogeneous parts 
as a semidirect product.
In conclusion, our superalgebra has the following
structure: 
$$
\widetilde{\sch}(n/\N)
=\left(\o(n)\times\osp(1/\N)\right)
\,\semidirectproduct\,
{\rm h}(n/\N),
\equation
$$
generalizing what Beckers et al.~[7] found for $N=N_+=2$.
In most frequent cases, when all $epsilon$'s are equal to $1$ and $N=N_+$, 
we denote our Schr\"odinger algebras simply by $\sch(n/N)$.

\goodbreak
\medskip

\kikezd{Discussion.}

Let us recall~[4] that the $\math{SL}(2,\bR)$ subgroup of the Schr\"odinger
group  actually shows up in the presymplectic framework over space-time as a 
covering group of the projective transformations of the time axis. 
Our $I$-type Ansatz, (4.4-5) amounts to decomposing $\bR^{2n}$ by viewing it as
a tensor product,
$$
\bR^{2n}\cong\bR^n\otimes\bR^2,
\equation
$$
and super-extending it by extending the `time-like' factor
$\bR^2$ into
$\bR^{2/N}$, so that the super-symplectic phase space becomes
$$
\bR^{2n/nN}\cong\bR^n\otimes\bR^{2/N}.
\equation
$$

The structure of the (unextended) $I$-type superalgebra
$\sch(n/N_+,N_-)$
becomes clear by writing its action on the super-symplectic space in the form 
$$
\delta(\bP\bQ\,\xib_1\ldots\xib_N)
=
A(\bP\bQ\,\xib_1\ldots\xib_N)+(\bP\bQ\,\xib_1\ldots\xib_N)B+C,
\equation
$$
where (cf. (4.6) for the notation)
$$
\left\{\eqalign{
&A\in \o(n)
\ccr
&B=\pmatrix{
a&c&\overline\theta\ccr
b&-a&-\overline\sigma\ccr
{\sigma}&{\theta}&R\cr}\in\osp(1/N_+,N_-)
\ccr
&C=(\bV\bW\,\bPsi_1\ldots\bPsi_N)\in\bR^{2n/nN}.
\cr}\right.
\equation
$$

We note that the $I$-type extension amounts hence to super-extending 
the conformal subalgebra $\math{sl}(2,\bR)\cong\osp(1/0)$ into
$\osp(1/N_+,N_-)$.
In other words, it consists of those automorphisms that respect the
`space-supertime' factorization (4.21) much in the same way as the
Schr\"odinger algebra $\sch(n)$ does for the `space-time' 
factorization (4.20).



\chapter{Schr\"odinger superalgebras: exotic ($IJ$-type) extensions.}

Now we try to find other supersymmetric extensions of the Schr\"odinger
algebra. Let us hence return to the conditions (4.3), and inquire 
under what circumstances the
second one, for example,
$$
\Sigma\overline\Sigma'+\Sigma'\overline\Sigma=\lambda I_n,
\equation
$$
($\lambda\in\bR$) can be satisfied by $\Sigma$'s of the form
$$
\Sigma=\left(\Sigma^1\,\ldots\,\Sigma^\nu\right),
\equation
$$
where
$\Sigma^j$ is in $L(\bR^n,\bR^n)$ for $j=1,\ldots,\nu$. Such a matrix 
$\Sigma^j$ can be
decomposed into its symmetric and antisymmetric parts
(with repect to the metric
$G_j=\epsilon_jI_n$),
and condition (5.1), applied to $\Sigma$ and $\Sigma'$ with
${\Sigma'}^k=\delta^{jk}I_n$, entails
that any $\Sigma^j$ is
necessarily of the form $\sigma^j I_n+\Omega^j$, where 
$\Omega^j+\overline{\Omega^j}=0$. Thus $\Omega^j$ is an orthogonal matrix. 
Now Eq.~(5.1) requires
$$
\left(\Omega^j\right)^2=\lambda^j I_n.
\equation
$$
Thus, in the generic case $\lambda^j\neq0$, $\Omega^j$ is invertible, so
that space has to be even dimensional.  Let $\{\Omega_K\}$ be a basis of the
subspace spanned by those orthogonal matrices which satisfy Eq.~(5.3). Then it
is easy to verify that all three relations in Eq.~(4.3) hold if
$$
\left\{\matrix{
\Sigma=\left(\Sigma^1\ldots\Sigma^\nu\right)\hfill
\ccr
\Theta=\left(\Theta^1\ldots\Theta^\nu\right),\hfill
\cr
}\right.
\where
\left\{\matrix{
\Sigma^j=\sigma^jI_n+\sigma_*^{jK}\Omega_K
\ccr
\Theta^j=\theta^jI_n+\theta_*^{jK}\Omega_K.
\cr
}\right.
\equation
$$

Then some algebra yields the crucial condition
$$
\sigma_*^{jK}\left(\Omega_KA-A\Omega_K\right)=0
\equation
$$
for all $A\in {\math o}(n)$.
Thus, if $n > 2$, the only possibility that wouldn't break
the ${\math o}(n)$-symmetry, is $\sigma_*^{jK}=0$. In
other words, for $n > 2$, the $I$-type extension considered in Section $4$ is
the only possibility consistent with the Ansatz (5.2).

If, however the {\it space is
two-dimensional}, the rotation group is
Abelian, and one can have a non-trivial $\Omega$, namely
$$
\Omega=\sigma_* J,
\equation
$$
$J$ being the 
generator of rotations in the plane, $J=\pmatrix{0&1\cr-1&0\cr}$. 

Let us hence examine the planar case $n=2$ in some more detail. Now the
Grassmann space is $m=2\nu$-dimensional. Then the matrix $D$ in (4.2) can be
decomposed into $2\times2$ blocks, 
$$
D=\pmatrix{
D_1^1 &D_1^2 &\ldots &\ldotp\ccr
D_2^1 &\ldotp & &\ldotp\ccr
\vdots & &\ddots &\vdots\ccr
\ldotp &\ldotp &\ldots &D_\nu^\nu\cr
}.
$$
Since $D$ is orthogonal, $D\in{\math o}(2\nu_+,2\nu_-)$, each $2\times2$ block
is further seen to be of the form $D_k^j=R_k^jI+S_k^jJ$ where $I=I_2$,
$(R_k^j)$ is in $o(\nu)$ (a shorthand for ${\math o}(\nu_+,\nu_-)$), and 
$(S_k^j)$ is a symmetric $\nu\times\nu$ matrix. Hence we have proved that, for 
$n=2$, the Ansatz (4.5) can be generalized (see (5.4,6)) to
$$
\left\{
\eqalign{
&\Sigma=(\sigma^1 I+\sigma_*^1 J\,\ldots\,\sigma^\nu I+\sigma_*^\nu J)
\ccr
&\Theta=(\theta^1 I+\theta_*^1 J\,\ldots\,\theta^\nu I+\theta_*^\nu J),
\cr}\right.
\equation
$$
yielding the `exotic' super-Schr\"odinger algebra which 
consists of those endomorphisms of $\bR^{6/2\nu}$ of the form
$$
Z=
\pmatrix{
A+aI & bI & 0 & \bV &-\Sigma^1 &\ldots &-\Sigma^\nu
\ccr
cI & A-aI & 0 & \bW &-\Theta^1 &\ldots &-\Theta^\nu
\ccr
\half\overline\bW & -\half\overline\bV & 0 & u &
-\half\overline{\bPsi}{}^1 & \ldots &-\half\overline{\bPsi}{}^\nu
\ccr
0 & 0 & 0 & 0 & 0 &\ldots&0\ccr
\overline{\Theta}_1 &-\overline{\Sigma}_1 & 0 &\bPsi_1 &
D^1_1 & \ldots & D^\nu_1
\ccr
\vdots &\vdots &\vdots &\vdots &\vdots &\ddots &\vdots
\ccr
\overline{\Theta}_\nu &-\overline{\Sigma}_\nu & 0 &\bPsi_\nu &
D^1_\nu &\ldots &D^\nu_\nu
\ccr
}
\equation
$$
where $A=\omega J$; $a,b,c,u\in\bR$;
$\bV,\bW\in\bR^2$;
$\Sigma^j=\sigma^jI+\sigma^j_*J$,
$\Theta^j=\theta^jI+\theta^j_*J$;
$\bPsi_j\in\bR^2_1$; the $2\times{2}$ block
matrices $D^j_k$ being given for $j,k=1,\ldots,\nu$ by
$$
D^j_k=R^j_kI+S^j_k J
\equation
$$ 
with $R+\overline{R}=0$, i.e. $R\in\o(\nu)$ and
$S=\overline{S}$.

The components of the supermoment map are, this time
$$
\eqalign{
\langle\mu,Z\rangle=
&-\cJ\omega +\cH{c}-\cD{a}-\cK{b}
-\overline\cG\,\bV +\overline\cP\,\bW+\cM{u}
\cr
&-\half{\cH}_{jk}R^{jk}+\half{\cL}_{jk}S^{jk}
+\cQ_j\theta^j +\cQ^*_j\theta_*^j 
-\cS_j\sigma^j -\cS^*_j\sigma_*^j
+\overline{\bXi}{}^j\bPsi_j
}
\equation
$$
with $j,k=1,\ldots,\nu$.
(Again, indices in (5.10) are raised and lowered by means of the metric
$g_{ij}~=~\epsilon_i~\delta_{ij}$.)
Explicitly, we get the remarkably symmetric formul\ae
$$
\left\{
\matrix{
\cJ&=
\bQ\times{\bP},\hfill
&\cQ_j&=
\bP\cdot\xib_j,\hfill
\ccr
\cH&=\half||\bP||^2,\hfill
&\cQ^*_j&=
\bP\times\xib_j,\hfill
\ccr
\cD&=
\bP\cdot{\bQ},\hfill
&\cS_j&=
\bQ\cdot\xib_j,\hfill
\ccr
\cK&=\half||\bQ||^2,\hfill
&\cS^*_j&=
\bQ\times\xib_j,\hfill
\ccr
\cG&=\bQ,\hfill
&\bXi_j&=\xib_j,\hfill
\ccr
\cP&=\bP,\hfill
&
\ccr
\cM&=1,\hfill
&
\ccr
\cH_{jk}&=\cH_{[jk]}=
\xib_j\cdot\xib_k,\hfill
&
\ccr
\cL_{jk}&=\cL_{(jk)}=
\xib_j\times\xib_k.\qquad
\hfill
&
\cr
}\right.
\equation
$$ 
Note that, 
in the plane, the cross product of two vectors is a scalar,
${\bf u}\times{\bf v}
=
\varepsilon_{ij}u^iv^j\equiv{\overline{\bf u}}\,J\,{\bf v}$, 
where $\varepsilon_{ij}$ is the totally antisymmetric symbol, 
$\varepsilon_{12}=1$.
Hence, 
$\xib\times\xib=\xi^1\xi^2-\xi^2\xi^1=2\xi^1\xi^2$. 

Thus,  we now have $N=2\nu$ `$\cQ$-type' and `$\cS$-type 
charges, half of them symmetric ($I$-type), the other half antisymmetric
($J$-type), but only $\nu$ `${\bf T}$-type' vector-charges $\bXi_j$.
Note also that we now have two types of `fermionic Hamiltonians' namely the
{symmetric} $\cL_{jk}$'s and the {antisymmetric} $\cH_{jk}$'s.

The super-commutators and super-Poisson brackets, calculated in the usual way,
are rather complicated. They are listed in the Appendix. The particular case
of $\nu=1$, $\epsilon_1=\epsilon_2=1$, which plays a significant r\^ole in
Chern-Simons theory, is worked out in Section 8.



\chapter{A free spinning particle.}

Let us first 
consider a free,
spin-$\half$ particle in ordinary space and construct its supersymmetry algebra 
directly.
At the pseudoclassical level 
[9] including spin amounts to adding an anticommuting $3$-vector,
$\zetab=(\zeta^a)$, turned into Pauli matrices upon 
quantization. 
In the first part of this Section we work in the quantized setting, i.e. 
with commutators.
The Hamiltonian $\cH$ is formally the same as for spin
$0$, $\cH=\bp^2/(2m$), 
and is readily seen to be the square of the
conserved supercharge 
$$
\matrix{
\cQ={\displaystyle{1\over\sqrt{m}}\,\bp\cdot\zetab}
\qquad\qquad\hfill
&\hbox{(helicity)}.
\cr
}
\equation
$$ 
The commutator of $\cQ$ with boosts, $\cB=m\br-\bp t$, yields another
conserved charge, $\bXi$, which turns out to be proportional to $\zetab$, 
$$
\bXi=\sqrt{m}\,\zetab\qquad\qquad\hbox{(spin)}.
\equation
$$

Being the sum of two conserved quantities,
$$
\matrix{
\cJ^{ab}=r^ap^b-p^ar^b-\zeta^a\zeta^b\qquad\hfill
&\hbox{(total angular momentum),}\hfill
\cr
}
\equation
$$
is also conserved. The bosonic generators $\cH, \cJ, \cP, \cB, \cM$ 
(see (6.10) below), and the fermionic generators $\cQ$ and $\bXi$ form 
the $15$-dimensional Galilei super-algebra of Gauntlett et al.~[5]. 
Commuting the helicity, $\cQ$, with the expansion, $\cK$,
yields a new charge, namely
$$
\cS=\sqrt{m}\,\left(\br-{\bp\over m}t\right)\cdot\zetab
\qquad\hfill
\hbox{(super-expansion).}
\equation
$$

Hence we find an $18$-dimensional, $N=1$ supersymmetric
extension of the Schr\"odinger algebra~[5], whose 
commutation relations correspond
to those of $\widetilde{\math{sch}}(3/1)$ in (4.19). 

The above result can easily be further extended, using our framework. 
(From now on, we work again at the pseudo-classical level, i.e. 
with Grassmann variables). 
Let us chose 
$N$ anticommuting $n$-vectors, $\zeta=(\zetab_1,\ldots,\zetab_N)$.
A free non-relativistic particle with mass $m$ and spin ${N/2}$
is described by the action~[9]
$$
{\cal A}
=
\int\!\bigg\{\bp\cdot\dot\br
+
\half\sum_{j=1}^N{\overline{\zetab}_j\dot{\zetab}_j}
-{\bp^2\over2m}\bigg\}\ dt,
\equation
$$
with associated equations of motion
$\dot\bp=0$,
$\dot\br=\bp/m$,
$\dot\zeta=0$.
The classical motions are straight lines with $\zeta=\const$.
The `space of motions' is hence globally parametrized by
$$
\bP={\bp\over m},
\qquad
\bQ=\br-{\bp\over m}t,
\qquad
\xib_j={\zetab_j\over\sqrt{m}}.
\equation
$$

In Souriau's language~[8] adapted to this (super)setting, 
this amounts to
working with the `pre-symplectic' two-form 
$$
\sigma=d\overline\bp\wedge d\br+
\half\sum_{j=1}^N{d\overline{\zetab}_j\wedge d\zetab_j}
-d\left({\bp^2\over2m}\right)\wedge dt
\equation
$$
on the `evolution space' 
$$
{\cal E}=
\left\{
\pmatrix{\bp\cr\br\cr t\cr\zeta\cr}\;
\bigg\vert\;
\bp,\br\in\bR^n;
\;t\in\bR;\;
\zeta=\pmatrix{\zetab_1\cr\vdots\cr\zetab_N\cr},
\;
\zetab_j\in\bR^{n}_1\right\}.
\equation
$$
The 2-form $\sigma$ in Eq.~(6.7) is readily seen 
to project to the space of motions as  
$$
m\Bigl(
d\overline\bP\wedge d\bQ
+
\half\sum_{j=1}^N{d\overline{\xib}_j\wedge d\xib_j}
\Bigr),
\equation
$$
i.e. $m$-times the super-symplectic form $\omega$ in 
Eq.~(2.3) with $\epsilon_j=1$. 
According to our general theory, the system admits an
$N=N_+$-super-Schr\"odinger symmetry given by Eq.~(4.6).
The associated conserved quantites are obtained therefore by inserting the
parameters $\bP,\bQ$ and $\xi$ from Eq.~(6.6) into Eq.~(4.10), and multiplying by
$m$ to yield
$$
\left\{
\matrix{
\cJ^{ab}
&=&
{\displaystyle
r^ap^b-p^ar^b-\sum_{j=1}^N{\zeta_j^a\zeta_j^b}
}
\qquad\hfill
&\hbox{(angular momentum)}\hfill
\cr
\cH
&=&
{\displaystyle
{\bp^2\over 2m}
}
\hfill
&\hbox{(energy)}\hfill
\ccr
\cD
&=&
{\displaystyle
\bp.\left(\br-{\bp\over m}t\right)
}
\hfill 
&\hbox{(dilatations)}\hfill
\ccr
\cK
&=&
{\displaystyle
\half m\left(\br-{\bp\over m}t\right)^2
}
\hfill
&\hbox{(expansions)}\hfill
\ccr
\cH_{ij}
&=&
{\displaystyle
\zetab_i\cdot\zetab_j
}
\hfill
&\hbox{(fermionic Hamiltonians)}
\hfill
\ccr
\cQ_j
&=&
{\displaystyle
{1\over\sqrt{m}}\,\bp\cdot\zetab_j
}
\hfill
&\hbox{(helicities)}\hfill
\ccr
\cS_j
&=&
{\displaystyle
\sqrt{m}\,\left(\br-{\bp\over m}t\right)\cdot\zetab_j
}
\hfill
&\hbox{(super-expansions)}\hfill 
\ccr
\cB
&=&
m\br-\bp t\hfill
&\hbox{(Galilean boosts)}\hfill
\ccr
\cP
&=&
\bp\hfill
&\hbox{(linear momentum)}\hfill
\ccr
\cM
&=&
m\hfill
&\hbox{(mass)}\hfill
\ccr
\bXi_j
&=&
\sqrt{m}\,\zetab_j\hfill
&\hbox{(spins)}\hfill 
\cr
}\right.
\equation
$$

This generalizes the superalgebra of Gauntlett et al.~[5] 
from $N=1$ to any $N$. (The
expression of the angular momentum is consistent with having 
spin-$\half{N}$).
Notice that for $N\geq 2$, one also gets an extra bosonic
$\o(N)$ generated by the $\cH_{ij}$'s 
---~whose conservation is, however, trivial.

If $N\geq2$, an extra term can be added to the Hamiltonian. For 
$N=2$, for example, we have two anticommuting Grassmann vectors 
$\zetab_j$, and we can consider the new Hamiltonian
$$
\cH_{\rm tot}=\cH_B+\cH_F
=
{\bp^2\over2m}-i\omega\,
\left(\zetab_+\cdot\zetab_--\zetab_-\cdot\zetab_+
\right)
\equation
$$
with $\omega=\const$. Since the new term commutes with all bosonic
variables, the motion in space is undisturbed.
However, the $\zetab_\pm=\half(\zetab_1\pm i\zetab_2)$ satisfy rather 
$\dot{\zetab}_\pm=\mp i\omega\,\zetab_\pm$,
so that
$
\zetab_\pm(t)=e^{\mp i\omega t}\zetab_{\pm}(0).
$
The projection to the space of motions is hence given as
$\bP=\bp/m$,$\bQ=\br-\bp t/m$, supplemented by
$$
\xib_\pm=e^{\pm i\omega t}{\zetab_{\pm}\over\sqrt{m}}.
\equation
$$
Then it is easy to check that one still gets the same 
symplectic form (6.9). 
In other words, the space of motions is unchanged by the extra 
term, and it has therefore the same $\widetilde{\sch}(n/N)$ supersymmetry as for 
$\omega=0$ --- realized in a less trivial way.

Let us mention that, if the space is two dimensional, $n=2$, we 
have, in addition to the $I$-type SUSY considered before, also the
`exotic' SUSY described in Section 5, with the usual bosonic 
generators $\cH,\cD,\cK$, $\nu(\nu-1)/2$ `fermionic Hamiltonians'
$\cH_{ij}=\zetab_i\cdot\zetab_j$,
but also 
$\nu(\nu+1)/2$ symmetric charges $\cL_{ij}=\zetab_i\times\zetab_j$. 
The angular momentum
is simply the scalar
$\cJ=\br\times\bp$.

The bosonic generators are supplemented by $\nu$ $I$-type and 
$\nu$ $J$-type supercharges,
as well as $\nu$ spin vectors $\bXi_j=\zetab_j/\sqrt{m}\in\bR_1^2$,
$j=1,\ldots,\nu$. The structure of this algebra is described, for 
$\nu=1$, in Section 8.



\chapter{The symmetries of the harmonic oscillator.}

Let us first rederive the Schr\"odinger symmetry for an $n$-dimensional bosonic 
harmonic oscillator~[6] in our framework.
Let us indeed consider the
oscillator with Hamiltonian $H_B=\bp^2/(2m)+m\omega^2\br^2/2$. The
classical trajectories are 
$
\br(t)={\bf A}\cos\omega t+(1/m\omega){\bf B}\sin\omega t.
$
The space of motions is therefore $\bR^{2n}$, parametrized by 
${\bf A}$ and ${\bf B}$. In
fact, using
$$
{\bf A}=\br\cos\omega t-{\bp\over m\omega}\sin\omega t
\and
{\bf B}=m\omega\br\sin\omega t+\bp\cos\omega t,
\equation
$$
it is readily verified that the presymplectic $2$-form $\sigma$
 of the evolution
space ${\cal E}=\bR^{2n+1}$ parametrized by the triples $(\br,\bp,t)$ 
 projects to
the space of motions ${\cal E}/\ker(\sigma)$ 
as the canonical symplectic structure of $\bR^{2n}$, 
$$
\sigma=d\overline\bp\wedge d\br-dH_B\wedge dt
=d\overline{\bf B}\wedge d{\bf A}
=m(d\overline\bP\wedge d\bQ)
\equation
$$
where
$$
\bP={{\bf B}\over m}
\and
\bQ={\bf A}.
\equation
$$
Thus, the space of motions is the {\it same} symplectic vector space as for a
free particle --- and carries therefore a Schr\"odinger symmetry~[6]. The
components of the moment map read 
$$
\left\{
\eqalign{
&\cJ_{ab}=r_ap_b-r_bp_a\quad\hfill
\ccr
&\cP={\bf B}=m\omega\br\sin{\omega t}+\bp\cos{\omega t}\hfill
\ccr
&\cB=m{\bf A}=m\br\cos{\omega t}-{\bp\over\omega}\sin{\omega t}\hfill
\ccr
&\cH={1\over2m}{\bf B}^2=
\half\bigg(m\omega^2\br^2\sin^2{\omega t}+{\bp^2\over m}
\cos^2{\omega t} +\omega\,\br\cdot\bp\sin2{\omega t}\bigg)\hfill
\ccr
&\cM=m
\ccr
&\cD={\bf B}\cdot{\bf A}=
\half\left(m\omega\br^2-{\bp^2\over m\omega}\right)\sin2{\omega t}
+\br\cdot\bp\cos2{\omega t}\hfill
\ccr
&\cK=\half m{\bf A}^2=\half m\bigg(\br^2\cos^2{\omega t}
+\left({\bp\over m\omega}\right)^2\sin^2{\omega t} 
-{\br\cdot \bp\over m\omega}\sin 2{\omega t}\bigg).\hfill
\ccr}
\right.
\equation
$$

The symmetry generators are 
combinations of those in Eq.~(4.10). For example, a time-translation for 
the oscillator, $\delta t=\epsilon$,
 appears, in `free particle' (i.e. space of motions) language, as
a time  translation by $\epsilon$, followed by an expansion by 
$\omega^2\epsilon$, etc. In the so-called oscillator representation, they
are expressed as
$$
\left\{
\eqalign{
J_{ab}&=\cJ_{ab}=r_ap_b-r_bp_a
\ccr
H_B&={\cH}+\omega^2\cK
={1\over2}\left({\bp^2\over m}+m\omega^2 \br^2\right)
\ccr
C_\pm&=\pm i\left(\cH-\omega^2\cK\pm i\omega\cD\right)
=\pm{i\over2m}e^{\mp2i\omega t}\left(\bp\pm im\omega \br\right)^2
\ccr
\bP_\pm&=\pm i\left(\cP\pm i\omega\cB\right)
=\pm ie^{\mp i\omega t}\left(\bp\pm im\omega \br\right)
\ccr
M&=\cM=m
\cr
}\right.
\equation
$$
with $a,b=1,\ldots,n$. Here $H_B$, $C_+$ and $C_-$ generate an
$\math{sl}(2,\bR)$ algebra; the angular momentum $J$ generates $\o(n)$;
finally, the $\bP_{\pm}$ and $M$ span an $n$-dimensional Heisenberg algebra 
$\rm{h}(n)$. 

The $N=2$ supersymmetric oscillator in $n$ space dimensions 
can be described by adding two anticommuting vectors, 
$\zetab_1$ and $\zetab_2$ 
(or $\zetab_\pm=\half\left(\zetab_1\pm i\zetab_2\right)$). The action is
$$
{\cal A}=\int\bp\cdot d\br
+\half\sum_{j=1}^2{\overline{\zetab}_jd{\zetab}_j}
-\left[{\bp^2\over2m}
+{m\omega^2\br^2\over2}
-i{\omega}\left(\zetab_+\cdot\zetab_--\zetab_-\cdot\zetab_+\right)
\right]dt.
\equation
$$

The equations of motions for $\bp$ and $\br$ 
are {\it identical} to those without spin, while for $\zeta$ we get
$d\zetab_1/dt=\omega\,\zetab_2$,
$d\zetab_2/dt=-\omega\,\zetab_1.$
It follows that the trajectory
$\br(t)$
is the same as above, 
and that
$\zetab_\pm(t)=e^{\mp i\omega t}\zetab_\pm(0).$
The evolution space is 
the same as for a free particle, see (6.8) with $N=2$, while
Souriau's two-form reads
$$
\sigma=d\overline\bp\wedge d\br
+\half\sum_{j=1}^2{d\overline{\zetab}_j\wedge d\zetab_j}
-d\left[{\bp^2\over2m}+{m\omega^2\br^2\over2}
-i{\omega}\left(\zetab_+\cdot\zetab_--\zetab_-\cdot\zetab_+\right)
\right]\wedge dt.
\equation
$$

The coordinates on the space of motions (now super-symplectomorphic to
$\bR^{2n/nN}$) can therefore be chosen as
$$
\bP={{\bf B}\over m},
\qquad 
\bQ={\bf A},
\qquad
\xib_\pm=e^{\pm i\omega t}\,{\zetab_\pm\over\sqrt{m}}.
\equation
$$

It is easy to see that $\sigma$ projects to the space of motions to yield once
again $m$-times the (super)symplectic form of Eq.~(2.3) with $N=2$ and
$\epsilon_j=1$. Inserting (7.8) into (4.10)  and multiplying by $m$ yields the
modified  expression of the angular momentum including a spin contribution, 
$$
\cJ^{ab}=r^ap^b-r^bp^a-\left(\zeta_+^a\zeta_-^b+\zeta_-^a\,\zeta_+^b\right),
\equation
$$ 
and provides us with further conserved quantities. 
Using 
$$\cP\pm im\omega\cB=e^{\mp i\omega t}\left(\bp\pm im\omega\br\right),
$$
we obtain
$$
\left\{\matrix{
\cQ_\pm\mp im\omega\cS_\pm\hfill
&=&\left(\bp
\mp im\omega\br\right)
\cdot
\displaystyle{\zetab_\pm\over\sqrt{m}},
\hfill
\cr\cr
\cQ_\pm\pm im\omega\cS_\pm\hfill
&=&e^{\mp2i\omega t}\left(\bp\pm im\omega\br\right)\cdot
\displaystyle{\zetab_\pm\over\sqrt{m}}\hfill
\cr
}\right.
\equation
$$
i.e. $Q_\pm$ and $S_\pm$ in Eq.~(1.3).
Similarly,
$$
H_F=-i\omega\cH_{12}=-i{\omega}\
\left(\zetab_+\cdot\zetab_--\zetab_-\cdot\zetab_+\right),
\equation
$$
coincides with the fermionic Hamiltonian in (1.2), and 
$$
\bXi_\pm=\sqrt{m}\,e^{\pm i\omega t}\zetab_\pm
\equation
$$
are the generators ${\bf T}_\pm$ in Eq.~(1.3). We have thus confirmed 
the $\widetilde{\math{sch}}(n/2)$ supersymmetry found by Beckers et al.
[7]. 

From Eqs (6.6) and (7.1,3) it follows also, that
$$
\left\{\eqalign{
t_{\rm free}&={1\over\omega}\,\tan(\omega t_{\rm osc}),
\ccr
{\br}_{\rm free}&={{\br}_{\rm osc}\over\cos(\omega t_{\rm osc})},
\ccr
\zetab_{\rm free}^\pm&=e^{\pm i(\omega t_{\rm osc}-
\tan(\omega t_{\rm osc}))}\zetab_{\rm osc}^{\pm}
\ccr}\right.
\equation
$$
extends Niederer's correspondence~[6] to the case of a free, spin-$1$ 
particle with Hamiltonian $H_{\rm tot}=H_B+H_F$, and the
supersymmetric oscillator.

We just mention, for completeness, that in $2$ space dimensions we can also have
exotic supersymmetry of Section 5. For $\nu=1$ 
(just one Grasmann vector $\zetab\in\bR^1_2$),
one gets the fermionic charges
$$
\eqalign{
&(\bp+im\omega\br)\cdot{\zetab\over\sqrt{m}},
\cr
&(\bp+im\omega\br)\times{\zetab\over\sqrt{m}},
\cr
&e^{2i\omega t}(\bp-im\omega\br)\cdot{\zetab\over\sqrt{m}},
\cr
&e^{2i\omega t}(\bp-im\omega\br)\times{\zetab\over\sqrt{m}}.
\cr}
\equation
$$
There is now no \lq $\cH_{12}$-type' extra bosonic charge. There is, 
however, an \lq $\cL_{12}$-type' charge, namely
$$
\cL={e^{2i\omega t}\over m}\,\zetab\times\zetab.
\equation
$$



\chapter{Chern-Simons-Matter Systems.}

In Ref.~[1] Leblanc, Lozano, and Min have constructed a
novel, $16$-dimensional superconformal extension of the
{\it planar} Galilei group. Describing their theory
goes beyond our scope here; we demonstrate, however, 
that their superalgebra is precisely our `exotic' extension 
described in Section~5.
Let us in fact assume that $n=2$, and consider the `$IJ$-type'
extension with $\nu=1$
and $\epsilon_1=\epsilon_2=1$. The 
Grassmann space is hence two-dimensional.
The corresponding `exotic super-Schr\"odinger algebra' is represented by
those matrices
$$
Z=
\pmatrix{
A+aI&bI&0&\bV&-{\sigma}I-{\sigma_{*}}J
\ccr
cI &A-aI&0&\bW&-{\theta}I-{\theta_{*}}J
\ccr
\half\overline\bW&-\half\overline\bV&0&u
&-\half\overline\bPsi
\ccr
0&0&0&0&0
\ccr
{\theta}I-{\theta_{*}}J&-{\sigma}I+{\sigma_{*}}J&0&\bPsi&S
\ccr
}
\equation
$$
where $I=I_2$, $J=J_2$;
$A=\omega J\in\o(2)$;
$a,b,c,u,r\in\bR$;
$\bV,\bW\in\bR^2$;
$S=rJ\in\o(2)$;
${\sigma},{\sigma_{*}},{\theta},{\theta_{*}}\in\bR_1$;
$\bPsi\in\bR^2_1$. 
(Note that there is no $R$ but there is an $S$ ---~cf. (5.9)).
The supercommutators $Z''=[Z,Z']$ are 
found as
$$
\left\{
\eqalign{
\omega''
&=
\sigma\theta'_*+\sigma'\theta_*-\sigma_*\theta'-\sigma'_*\theta
\cr
a''
&=
bc'-b'c-(\sigma\theta'+\sigma'\theta+\sigma_*\theta'_*+\sigma'_*\theta_*)
\cr
b''
&=
2(ab'-a'b)+2(\sigma\sigma'+\sigma_*\sigma'_*)
\cr
c''
&=
2(ca'-c'a)-2(\theta\theta'+\theta_*\theta'_*)
\cr
\bV''
&=
J(\omega \bV'-\omega'\bV)+a\bV'-a'\bV+b\bW'-b'\bW-(\sigma\bPsi'+\sigma'\bPsi)
\cr
&\qquad
-J({\sigma_*}\bPsi'+\sigma'_*\bPsi)
\cr
\bW''
&=
J(\omega \bW'-\omega'\bW)-a\bW'+a'\bW+c\bV'-c'\bV-(\theta\bPsi'+\theta'\bPsi)
\cr
&\qquad
-J(\theta_*\bPsi'+\theta'_*\bPsi)\hfill
\cr
u''
&=
-\overline\bV\,\bW'+\overline{\bV'}\,\bW -\overline\bPsi\,\bPsi'
\cr
r''
&=
2(\sigma\theta'_*+\sigma'\theta_*-\theta\sigma'_*-\theta'\sigma_*)
\cr
}\right.
\equation
$$
and
$$
\left\{\eqalign{
\sigma''
&=
\sigma_*\omega'-\sigma'_*\omega
+a\sigma'-a'\sigma+b\theta'-b'\theta
+r\sigma'_*-r'\sigma_*
\cr
\sigma''_*
&=
-\sigma\omega'+\sigma'\omega
+a\sigma'_*-a'\sigma_*+b\theta'_*-b'\theta_*
-r\sigma'+r'\sigma
\cr
\theta''
&=
-a\theta'+a'\theta
+\theta_*\omega'-\theta'_*\omega
-\theta_*r'+\theta'_*r
+c\sigma'-c'\sigma
\cr
\theta''_*
&=
-a\theta'_*+a'\theta_*
-\theta\omega'+\theta'\omega
+\theta r'-\theta' r
+c\sigma'_*-c'\sigma_*
\cr
\bPsi''
&=
-\sigma \bW'+\sigma' \bW
+\theta\bV'-\theta'\bV
\cr
&\;
-J(\theta_*\bV'-\theta'_*\bV)+J(\sigma_*\bW'-\sigma'_*\bW)
+J(r\bPsi'-r'\bPsi).
\cr}\right.
\equation
$$

The formula (5.11) of the 
supermoment map yields now $16$ supercharges, namely
$$
\left\{
\matrix{
\cJ\hfill 
&=&\bQ\times{\bP},\qquad
\qquad\hfill
&\cQ\hfill &=&{\bP}\cdot\xib,\hfill
\ccr
\cH\hfill &=&\half ||\bP||^2,\hfill
&\cQ^*\hfill &=&{\bP}\times\xib,\hfill
\ccr
\cD\hfill &=&\bP\cdot\bQ,\hfill
&\cS\hfill &=&\bQ\cdot\xib,\hfill
\ccr
\cK\hfill &=&\half ||\bQ||^2,\hfill
&\cS^*\hfill &=&\bQ\times\xib,\hfill
\ccr
\cB\hfill &=&\bQ,\hfill
&\bXi\hfill&=&\xib,\hfill
\ccr
\cP\hfill &=&\bP,\hfill& 
\ccr
\cM\hfill &=&1,\hfill
&\ccr
\cL&=&\xib\times\xib.\hfill&
\cr
}\right.
\equation
$$

Observe that there is just one super-translation vector, $\bPsi$, 
and hence a {\it single} `spin' vector $\bXi$, which does not 
contribute now to the angular momentum:  the space rotations do not affect the
Grassmann variable $\xib$. Thus $\cJ$ is merely the  orbital angular momentum.

The even Poisson brackets are those of the direct product of the
Schr\"odinger algebra $\sch(2)$  with an extra $\o(2)$.
Both sets of supercharges $\cQ,\cS$ and $\cQ^*,\cS^*$ extend the bosonic 
$\math{sl}(2,\bR)$ generated by $\cH,\cD,\cK$ into $\osp(1/1)$ but they 
fail to close with $\cL$ into $\osp(1/2)$, because the mixed
commutators $[\cQ,\cS^*]$ and $[\cQ^*,\cS]$ bring in the angular momentum:
$$
\matrix{
[\cQ,\cS]=0,\hfill
&[\cQ^*,\cS^*]=0,\hfill
\ccr
[\cQ,\cS^*]=-\cJ+\cL,\qquad\hfill
&[\cQ^*,\cS]=\cJ-\cL.\hfill
\cr}
\equation
$$
But $\cJ$ satisfies now non-trivial
commutation relations with the supercharges,
$$\matrix{
[\cJ,\cQ]=\cQ^*,\qquad\hfill
&[\cJ,\cQ^*]=-\cQ,\hfill
\ccr
[\cJ,\cS]=\cS^*,\hfill
&[\cJ,\cS^*]=-\cS.\hfill
\cr
}
\equation
$$
Thus, defining 
$$
\cY=\cJ-\cL=\bQ\times\bP-\xib\times\xib,
\equation
$$
the generators $\cH,\cD,\cK,\cY$ and 
$\cQ,\cQ^*,\cS,\cS^*$ satisfy once more the $\osp(1/2)$ relations,
$$
\matrix{
[\cQ,\cD]\hfill&=&\cQ\hfill
&[\cQ^*,\cD]&=&\cQ^*
\ccr
[\cQ,\cK]\hfill&=&\cS\hfill
&[\cQ^*,\cK]\hfill&=&\cS^*\hfill
\ccr
[\cQ,\cH]\hfill&=&0,\hfill
&[\cQ^*,\cH]\hfill&=&0\hfill
\ccr
[\cQ,\cY]\hfill&=&\cQ^*\hfill
&[\cQ^*,\cY]\hfill&=&-\cQ\hfill
\ccr
[\cS,\cD]\hfill&=&-\cS\hfill
&[\cS^*,\cD]\hfill&=&-\cS^*\hfill
\ccr
[\cS,\cK]\hfill&=&0\hfill
&[\cS^*,\cK]\hfill&=&0\hfill
\ccr
[\cS,\cH]\hfill&=&-\cQ\hfill
&[\cS^*,\cH]\hfill&=&-\cQ^*\hfill
\ccr
[\cS,\cY]\hfill&=&\cS^*\hfill
&[\cS^*,\cY]\hfill&=&-\cS\hfill
\ccr
[\cQ,\cQ]\hfill&=&-2\cH\qquad\qquad\hfill
&[\cQ^*,\cQ^*]\hfill&=&-2\cH\hfill
\ccr
[\cS,\cS]\hfill&=&-2\cK\hfill
&[\cS^*,\cS^*]\hfill&=&-2\cK\hfill
\ccr
[\cQ,\cQ^*]&=&0\hfill
&[\cS,\cS^*]\hfill&=&0\hfill
\ccr
[\cQ,\cS]\hfill&=&-\cD\hfill
&[\cQ^*,\cS^*]\hfill&=&-\cD\hfill
\ccr
[\cQ,\cS^*]\hfill&=&-\cY\hfill
&[\cQ^*,\cS]\hfill&=&\cY.\hfill
\cr}
\equation
$$

On the other hand,
$$
\cZ=\cJ-\half\,\cL=\cQ\times\cP-\half\xib\times\xib
\equation
$$
commutes with all generators of $\osp(1/2)$, so that the 
homogeneous part is once more a direct product, $\o(2)\times\osp(1/2)$.

The generator $\cZ$ plays the r\^ole of ordinary rotations, and
 $\cY$ behaves as a fermionic Hamiltonian in the conventional 
case.
Both mix spatial and internal 
rotations. Since $\cZ$ satisfies the same commutation relations with the 
bosonic generators as $\cJ$ does in (4.13), $\cZ, \cH,\cD,\cK, 
\cM,\cB,\cP$ form the standard Schr\"odinger algebra $\widetilde{\sch}(2)$.

The generators $\cB,\cP$ and $\bXi$ 
span the super-Heisenberg algebra $\rm{h}(2/1)$,
$$
[\cB^a,\Xi^b]=0,\qquad
[\cP^a,\Xi^b]=0,\qquad
[\Xi^a,\Xi^b]=-\cM\,\delta^{ab}.
\equation
$$

The remaining relations
are exactly the same as in the `$I$-type' case, showing that the structure 
of our new algebra is 
$$
\widetilde{\sch}_e(1)\cong\left(\o(2)\times\osp(1/2)\right)
\,\semidirectproduct\,\rm{h}(2/1),
\equation
$$
cf. Eq.~(4.19). The only difference is that there is now just one 
super-translation.

Finally, the commutation relations of the super\-algebra of Leblanc
et al.~[1] are indeed recovered from those (8.5-10) of our exotic
super-Schr\"odinger algebra $\widetilde{\sch}_e(1)$ by setting
$$
\matrix{
&\hbox{\underbar{LLM}}\qquad\qquad\hfill&&\qquad\hbox{\underbar{DH}}
\hfill\ccr
&J\hfill&=&\qquad\cJ+\cM-{\smallover1/4}\cL\hfill
\ccr
&H\hfill&=&\qquad\cH\hfill
\ccr
&K\hfill&=&\qquad\cK\hfill
\ccr
&D\hfill&=&\qquad\half\cD\hfill
\ccr
&G_\pm\hfill&=&\qquad\mp\cB_1+i\cB_2\hfill
\ccr
&P_\pm\hfill&=&\qquad\mp\cP_1+i\cP_2\hfill
\ccr
&N_F\hfill&=&\qquad-\half\cL\hfill
\ccr
&N_B\hfill&=&\qquad\cM+\half\cL\hfill
\ccr
&Q_1\hfill&=&\qquad i\Xi^1-\Xi^2\hfill
\ccr
&Q_1^*\hfill&=&\qquad\Xi^1-i\Xi^2\hfill
\ccr
&Q_2\hfill&=&\qquad\half(i\cQ-\cQ^*)\hfill
\ccr
&Q_2^*\hfill&=&\qquad\half(\cQ-i\cQ^*)\hfill
\ccr
&F\hfill&=&\qquad\half(\cS+i\cS^*)\;\hfill
\ccr
&F^*\hfill&=&\qquad\half(i\cS+\cS^*).\hfill
\ccr
}
$$



\chapter{Supersymmetry of the monopole and of the magnetic vortex.}

A few years ago, Jackiw~[11] pointed out that a spin-$0$ particle in a Dirac
monopole field has an $\o(2,1)$ dynamical symmetry, generated by the
spin-$0$ Hamiltonian, $\widehat\cH_0=\pib^2/(2m)$,
by the dilatation and by the expansion,
$$
\widehat\cD
=
-2t\widehat\cH_0
+
\half\left(\pib\cdot\br+\br\cdot\pib\right)
\and
\widehat\cK
=
t^2\widehat\cH_0
-
t\widehat\cD
+
\half m\br^2,
\equation
$$ 
where $\pib=\bp-e{\bf A}$ is associated with a monopole vector
potential, ${\bf A}$.
This result was extended to spin-$\half$ particles by D'Hoker and
Vinet~[12] who have shown that for the Pauli Hamiltonian
$$
\widehat\cH
=
{1\over2m}\left(\pib^2-e{\bf B}\cdot\sigmab\right),
\equation
$$
not only the conformal generators $\widehat\cD$ and $\widehat\cK$,
but also the `fermionic' operators
$$
\widehat\cQ
=
{1\over\sqrt{2m}}\,\pib\cdot\sigmab
\and
\widehat\cS
=
\sqrt{m\over2}\,\br\cdot\sigmab-t\widehat\cQ
\equation
$$
are conserved.
Thus, adding the total angular momentum, the spin system admits an
$\o(3)\times\osp(1/1)$ conformal supersymmetry.

Recently, Jackiw~[13] found that the $\o(2,1)$ symmetry,
generated by $\widehat\cD$ and $\widehat\cK$ is also present for a magnetic
vortex; it combines with the angular momentum and the `exotic' $N=2$
supersymmetry~[14] into an $\o(2)\times\osp(1/2)$
superalgebra~[15].

We first show this in a quantum-mechanical context.

Let us start with a spin-$\half$ particle in a static magnetic field
${\bf B}=B(x,y)\hat{\bf z}$.
Dropping the irrelevant $z$ variable, we work in the plane.
The model is described by the Pauli Hamiltonian
(9.2) with $B=\math{rot}\,{\bf A}\equiv\epsilon^{ij}\partial_iA_j$.
It is easy to see that $\widehat\cH$ is a perfect
square in two different ways:
both operators
$$
\widehat\cQ
=
{1\over\sqrt{2m}}\,\pib\cdot\sigmab
\and
\widehat\cQ^*
=
{1\over\sqrt{2m}}\,\pib\times\sigmab,
\equation
$$
where $\sigmab=(\sigma_1,\sigma_2)$, satisfy the anticommutation relations
$$
[\widehat\cQ,\widehat\cQ]_+=[\widehat\cQ^*,\widehat\cQ^*]_+=2\widehat\cH.
\equation
$$
Thus, for any static, purely magnetic field in the plane, $\widehat\cH$ is an
$N\=2$ supersymmetric Hamiltonian~[14].  

Let us assume henceforth that $B$ is the field of a point-like
magnetic vortex directed along the $z$-axis, $B=\Phi\,\delta(\br)$,
where $\Phi$ is the total magnetic flux.
It is  straightforward to check that $\widehat\cD$ and $\widehat\cK$
as in Eq.~(9.1) generate, along with $\widehat\cH$,
the $\o(2,1)$ Lie algebra, to which the angular momentum,
$\widehat\cJ=\br\times\pib$, adds an extra $\o(2)$.
Commuting $\widehat\cQ$ and $\widehat\cQ^*$ with the expansion,
$\widehat\cK$, yields two more fermionic generators, namely
$$
\left\{\eqalign{
\widehat\cS
&=
i[\widehat\cQ,\widehat\cK]
=\sqrt{m\over2}\left(\br-{\pib\over m}t\right)\cdot\sigmab,
\ccr
\widehat\cS^*
&=
i[\widehat\cQ^*,\widehat\cK]
=
\sqrt{m\over2}\left(\br-{\pib\over m}t\right)\times\sigmab.
\cr}\right.
\equation
$$

Then the same calculation as in Section~8 shows that
$\widehat\cY=\br\times\pib+\sigma_3,\widehat\cH,\widehat\cD,\cK$ and
$\widehat\cQ,\widehat\cQ^*,\widehat\cS,\widehat\cS^*$ 
span the $\osp(1/2)$ superalgebra.
Adding also $\widehat\cZ=\br\times\pib+\half\sigma_3$, which
commutes with all generators of $\osp(1/2)$, we conclude that 
the full symmetry superalgebra is the direct product
$\osp(1/2)\times\o(2)$, generated by
$$
\left\{
\matrix{
\widehat\cY\hfill
&=&
\br\times\pib+\sigma_3,
\qquad\qquad\hfill
&
\widehat\cQ\hfill
&=&
\displaystyle{1\over\sqrt{2m}}\,\pib\cdot\sigmab,\hfill
\ccr
\widehat\cH\hfill
&=&
\displaystyle{1\over2m}\,
\left(\pib^2-eB\sigma_3\right),
\qquad\quad\hfill
&
\widehat\cQ^*\hfill
&=&
\displaystyle{1\over\sqrt{2m}}\,\pib\times\sigmab,\hfill
\ccr
\widehat\cD\hfill
&=&
\half\,(\pib\cdot\bq+\bq\cdot\pib)
+t\displaystyle{eB\over m}\,\sigma_3,
\quad\hfill
&
\widehat\cS\hfill
&=&
\sqrt{\displaystyle{m\over2}}\,\bq\cdot\sigmab,\hfill
\ccr
\widehat\cK\hfill
&=&
\half m\bq^2,\hfill
&
\widehat\cS^*\hfill
&=&
\sqrt{\displaystyle{m\over2}}\,\bq\times\sigmab,\hfill
\ccr
\widehat\cZ\hfill
&=&
\br\times\pib+\half\sigma_3,\hfill
&
\ccr
}\right.
\equation
$$
where we have introduced
$\bq\equiv\br-t\pib/m.$

Now we re-derive these results in our framework.

We shall be concerned with the supersymplectic space $\bR^{2n/n}$, 
where $n=2,3$ to deal respectively with the case of the magnetic vortex 
$F=\half F_{jk}\,dr^j\wedge{dr^k}=\Phi\delta(\br)\,dx\wedge{dy}$,
and the Dirac monopole
$F={1\over2}\epsilon_{ijk}B^i\,dr^j\wedge dr^k$ with
${\bf B}=g\br/r^3$.
The {\it minimal coupling} prescription amounts to replacing
the standard supersymplectic structure of $\bR^{2n/n}$,
parametrized by $(\br,\pib,\xib)$, see (2.3) by
$$
\omega
=
d\pi_j\wedge dr^j + \half d\xi_j\wedge d\xi^j 
+\half eF_{jk}\,dr^j\wedge dr^k
\equation
$$
where $e$ is the electric charge of the test particle. 

\goodbreak

\kikezd{1) The Dirac monopole.}

The Pauli Hamiltonian,
$$
\cH={1\over2m}\left(\pib^2 + eF_{ij}\xi^i\xi^j\right),
\equation
$$
gives rise to the Hamiltonian vectorfield $X_{\cH}$
(according to $\imath(X_{\cH})\omega=-d\cH$), 
$$
mX_\cH
=
\pi^j\partial_{r^j} 
-e\left(
F_{ij}\pi^i + \half\partial_jF_{k\ell}\xi^k\xi^\ell
\right)
\partial_{\pi_j}
-eF_{ij}\xi^i\partial_{\xi_j}.
\equation
$$

Routine calculation using the homogeneity property
$r^j\partial_jF_{k\ell}=-2F_{k\ell}$,
shows furthermore that $\cH$ and
$$\left\{
\matrix{
\cD=\pib\cdot\br-2t\cH\hfill
&
\hbox{with}\hfill
&
X_\cD=r^j\partial_{r^j}-\pi_j\partial_{\pi_j}-tX_\cH,\hfill
\ccr
\cr
\cK
=
\half m\br^2-t\cD+t^2\cH\hfill
&
\hbox{with}\hfill
&
X_\cK=-r_j\partial_{\pi_j}-tX_\cD+t^2X_\cH
\hfill
\cr}\right.
\equation
$$
form, under Poisson brackets $\{f,g\}=X_fg$,
and for each value of $t$,
an algebra isomorphic to $\o(2,1)$.
Moreover, the superfunctions
$$
\cQ={\pib\cdot\xib\over\sqrt{m}}
\and
\cS=\sqrt{m}\,\br\cdot\xib-t\cQ,
\equation
$$
whose hamiltonian vector fields read
$$
X_\cQ
=
{1\over\sqrt{m}}\left(
\xi^j\partial_{r^j}
-eF_{ij}\xi^i\partial_{\pi_j}
-\pi^j\partial_{\xi^j}\right)
\equation
$$
and
$$
X_\cS
=
-\sqrt{m}\left(
\xi_j\partial_{\pi_j}
+r^j\partial_{\xi^j}\right)
-
tX_\cQ,
\equation
$$
extend the bosonic symmetry algebra into $\osp(1/1)$.

The extra rotation generators read
$$
X_{\cZ_{jk}}
=
r_j\partial_{r^k}-r_k\partial_{r^j}
+\pi_j\partial_{\pi^k}-\pi_k\partial_{\pi^j}
+\xi_j\partial_{\xi^k}-\xi_k\partial_{\xi^j}
\equation
$$
with
$$
\cZ_{jk}=2r_{[j}\pi_{k]}-\xi_j\xi_k.
\equation
$$

\goodbreak

\kikezd{2) The magnetic vortex.}

The vortex case is quite similar: using 
$x\,\delta(\br)=y\,\delta(\br)=0$
and
$\br\cdot\delta'(\br)=-2\delta(\br)$,
it is straightforward to find nine Hamiltonian vector fields
$X_\cH, X_\cD, X_\cK, X_\cY, X_\cQ, X_{\cQ^*}, X_\cS, X_{\cS^*}$
and $X_\cZ$ which leave the symplectic form (9.8) invariant, 
and to check that their Poisson brackets satisfy the
$\o(2)\times\osp(1/2)$ supercommutation relations.
One readily checks that
$$
\left\{
\matrix{
\cY=\br\times\pib-\xib\times\xib,
\hfill
&
\cQ=\displaystyle{\pib\cdot\xib\over\sqrt{m}},
\hfill\ccr
\cH=\displaystyle{1\over2m}\left(\pib^2+eB\xib\times\xib\right),
\qquad\hfill
&
\cQ^*=\displaystyle{\pib\times\xib\over\displaystyle\sqrt{m}},
\hfill\ccr
\cD=\pib\cdot\br-2t\cH,
\hfill
&
\cS=\displaystyle{\sqrt{m}\,\br\cdot\xib-t\cQ},
\hfill\ccr
\noalign{\medskip}
\cK=\half m\br^2-t\cD+t^2\cH,
\hfill
&\cS^*=\displaystyle{\sqrt{m}\,\br\times\xib-t\cQ^*},
\hfill\ccr
\noalign{\medskip}
\cZ=\displaystyle{\br\times\pib-\half\xib\times\xib},
\hfill&\ccr
}\right.
\equation
$$
actually span the pseudoclassical analog of the operator 
superalgebra (9.7).

\bigskip

\parag
{\bf Acknowledgements.}
We are indebted to Professors J.-G.~Demers, M.~Leblanc
and P.~Townsend for helpful correspondence.

\vfill\eject


\centerline{\bf Appendix}

\def\qqhf{\qquad\qquad\hfill}
\def\qhf{\qquad\hfill}

In two configuration-space dimensions, $n=2$,
the $IJ$-type Ansatz (5.7), yields the following `exotic'
superalgebra (see (5.10,11)):
the even super-Poisson brackets read
$$
\matrix{
[\cJ,\cG^a]&=-J^a_b\cG^b\qqhf
[\cJ,\cP^a]&=-J^a_b\cP^b\qqhf
\ccr
[\cJ,\cQ_j]&=\cQ^*_j\qqhf
[\cJ,\cQ^*_j]&=-\cQ_j\qqhf
\ccr
[\cJ,\cS_j]&=\cS^*_j\qqhf
[\cJ,\cS^*_j]&=-\cS_j\qqhf
\ccr
[\cH,\cD]&=2\cH\qqhf
[\cH,\cK]&=\cD\qqhf
\ccr
[\cH,\cG^a]&=\cP^a\qqhf
&\qhf
\ccr
[\cH,\cS_j]&=\cQ_j\qqhf
[\cH,\cS^*_j]&=\cQ^*_j\qqhf
\ccr
[\cD,\cK]&=2\cK\qqhf
[\cD,\cG^a]&=\cG^a\qqhf
\ccr
[\cD,\cP^a]&=-\cP^a\qqhf
&\qhf
\ccr
[\cD,\cQ_j]&=-\cQ_j\qqhf
[\cD,\cQ^*_j]&=-\cQ^*_j\qqhf
\ccr
[\cD,\cS_j]&=\cS_j\qqhf
[\cD,\cS^*_j]&=\cS^*_j\qqhf
\ccr
[\cK,\cP^a]&=-\cG^a\qqhf
&\qhf
\ccr
[\cK,\cQ_j]&=-\cS_j\qqhf
[\cK,\cQ^*_j]&=-\cS^*_j\qqhf
\ccr
[\cG^a,\cP_b]&=-\cM\delta^a_b\qqhf
&\qhf
\ccr
[\cG^a,\cQ_j]&=-\Xi^a_j\qqhf
[\cG^a,\cQ^*_j]&=-J^a_b\Xi^b_j\qqhf
\ccr
[\cP^a,\cS_j]&=\Xi^a_j\qqhf
[\cP^a,\cS^*_j]&=J^a_b\Xi^b_j,\qqhf
\ccr
}
\leqno(A.1)
$$
$$
\matrix{
[\cH_{ij},\cH_{k\ell}]&=
\cH_{ik}g_{j\ell}-\cH_{jk}g_{i\ell}
+\cH_{j\ell}g_{ik}-\cH_{i\ell}g_{jk}\qqhf
&\qhf
\ccr
[\cH_{ij},\cL_{k\ell}]&=
-\cL_{ik}g_{j\ell}+\cL_{jk}g_{i\ell}
+\cL_{j\ell}g_{ik}-\cL_{i\ell}g_{jk},\qqhf
&\qhf
\ccr
}
\leqno(A.2)
$$
$$
\matrix{
[\cH_{jk},\cQ_{\ell}]&=-2\cQ_{[j}g_{k]\ell}\qqhf
[\cH_{jk},\cQ^*_{\ell}]&=-2{\cQ^*}_{[j}g_{k]\ell}\qqhf
\ccr
[\cH_{jk},\cS_{\ell}]&=-2\cS_{[j}g_{k]\ell}\qqhf
[\cH_{jk},\cS^*_{\ell}]&=-2{\cS^*}_{[j}g_{k]\ell}\qqhf
\ccr
[\cH_{jk},\Xi^a_{\ell}]&=-2{\Xi^a}_{[j}g_{k]\ell},\qqhf
&\qhf
\ccr
}
\leqno(A.3)
$$
$$
\matrix{
[\cL_{ij},\cL_{k\ell}]&=
\cH_{ik}g_{j\ell}+\cH_{jk}g_{i\ell}
+\cH_{j\ell}g_{ik}+\cH_{i\ell}g_{jk},
&\qqhf
\ccr}
\leqno(A.4)
$$
$$
\matrix{
[\cL_{jk},\cQ_{\ell}]&=2{\cQ^*}_{(j}g_{k)\ell}\qqhf
[\cL_{jk},\cQ^*_{\ell}]&=-2\cQ_{(j}g_{k)\ell}\qqhf
\ccr
[\cL_{jk},\cS_{\ell}]&=2{\cS^*}_{(j}g_{k)\ell}\qqhf
[\cL_{jk},\cS^*_{\ell}]&=-2\cS_{(j}g_{k)\ell}\qqhf
\ccr
[\cL_{jk},\Xi^a_{\ell}]&=2J^a_b{\Xi^b}_{(j}g_{k)\ell},\qqhf
&\qhf
\ccr
}
\leqno(A.5)
$$
as for the odd super-Poisson brackets, they retain the form
$$
\matrix{
[\cQ_j,\cQ_k]&=-2\cH\,g_{jk}\qqhf
&\qhf
\ccr
[\cQ_j,\cS_k]&=-\cD\,g_{jk}+\cH_{jk}\qqhf
[\cQ_j,\cS^*_k]&=-\cJ\,g_{jk}+\cL_{jk}\qqhf
\ccr
[\cQ_j,\Xi^a_k]&=-\cP^a\,g_{jk}\qqhf
&\qhf
\ccr
[\cQ^*_j,\cQ^*_k]&=-2\cH\,g_{jk}\qqhf
&\qhf
\ccr
[\cQ^*_j,\cS_k]&=\cJ\,g_{jk}-\cL_{jk}\qqhf
[\cQ^*_j,\cS^*_k]&=-\cD\,g_{jk}+\cL_{jk}\qqhf
\ccr
[\cQ^*_j,\Xi^a_k]&=J^a_b\cP^b\,g_{jk}\qqhf
&\qhf
\ccr
[\cS_j,\cS_k]&=-2\cK\,g_{jk}
\qqhf
[\cS_j,\Xi^a_k]&=-\cG^a\,g_{jk}\qqhf
\ccr
[\cS^*_j,\cS^*_k]&=-2\cK\,g_{jk}\qqhf
[\cS^*_j,\Xi^a_k]&=J^a_b\cG^b\,g_{jk}\qqhf
\ccr
[\Xi^a_j,\Xi^b_k]&=-\cM\,\delta^{ab}\,g_{jk}.\qqhf
&\qhf
\ccr
}
\leqno(A.6)
$$

\vfill\eject


\centerline{\bf References}

\reference
M.~Leblanc, G.~Lozano and H.~Min,
{\it Extended superconformal Galilean symmetry in Chern-Simons matter systems},
Ann. Phys. (N.Y.) {\bf 219}, 328 (1992).

\reference
R.~Jackiw and So-Young Pi, 
{\it Classical and quantal nonrelativistic Chern-Simons theory},
Phys. Rev. {\bf D42}, 3500 (1990);
G.~V.~Dunne, R.~Jackiw, So-Young Pi and C.~A.~Trugenberger,
{\it Self-dual Chern-Simons solitons and two-dimensional non-linear
equations}, 
Phys. Rev. {\bf D43}, 1332 (1991).

\reference
U.~Niederer, 
{\it The maximal kinematical 
invariance group of the free Schr\"odinger equation},
Helv. Phys. Acta {\bf 45}, 802 (1972);
C.~R.~Hagen, 
{\it Scale and conformal transformations in Galilean-covariant
field theory},
Phys. Rev. {\bf D5}, 377 (1972);
G.~Burdet, M.~Perrin, 
{\it Many-body realization of the Schr\"odinger algebra},
Lett. Nuovo Cimento {\bf 4}, 651 (1972).

\reference
C.~Duval, G.~Burdet, H.~P.~K\"unzle and M.~Perrin,
{\it Bargmann structures and Newton-Cartan theory},
Phys. Rev. {\bf D31}, 1841 (1985);
W.~M.~Tulczyjew, 
{\it An intrinsic formulation of nonrelativistic analytical mechanics
and wave mechanics},
J. Geom. Phys. {\bf 2}, No. 3, 93 (1985);
V.~Husssin and S.~Sinzinkayo, 
{\it Conformal symmetry and constants of motion},
J. Math. Phys. {\bf 26}, 1072 (1985); 
V.~Hussin and M.~Jacques, 
{\it On non-relativistic conformal symmetries and invariant tensor fields},
J. Phys. {\bf A19}, 3471 (1986);
M.~Omote, S.~Kamefuchi, Y.~Takahashi, Y.~Ohnuki,
{\it Galilean symmetries},
in {\it Symmetries in Science III}, Proc '88 Schloss Hofen Meeting,
Gruber and Iachello (eds), p. 323 Plenum: N.Y. (1989);
M.~Henkel, in {\it Finite-size scaling and numerical simulation of
statistical systems}, Ch. VIII, p. 353, ed. V.~Privman, World Sci. (1990);
C.~Duval, G.~Gibbons and P.~Horv\'athy,
{\it Celestial mechanics, conformal structures, and gravitational waves},
Phys. Rev. {\bf D43}, 3907 (1991).

\reference
J.~P.~Gauntlett, J.~Gomis and P.~K.~Townsend,
{\it Supersymmetry and the physical-phase-space formulation of 
spinning particles}, Phys. Lett. {\bf B248}, 288 (1990).

\reference
U.~Niederer,
{\it The maximal kinematical 
invariance group of the harmonic oscillator},
Helv. Phys. Acta {\bf 46}, 192 (1973).


\reference
J.~Beckers and V.~Hussin,
{\it Dynamical supersymmetries of the harmonic oscillator},
Phys. Lett. {\bf A118}, 319 (1986);
J.~Beckers, D.~Dehin and V.~Hussin,
{\it Symmetries and supersymmetries of the quantum harmonic oscillator},
J. Phys. {\bf A20}, 1137 (1987).

\reference
J-M Souriau, {\it Structure des syst\`emes dynamiques}, Dunod: Paris (1969).

\reference
F.~A.~Berezin and M.~S.~Marinov, 
{\it Classical spin and Grassmann algebra},
JETP Lett. {\bf 21}, 321 (1975);
{\it Particle spin dynamics as the Grassmann variant of classical mechanics}, 
Ann. Phys. (N.Y.) {\bf 104}, p. 336, (1977)
F.~A.~Berezin, {\it Introduction to superanalysis}, Ed. by A.~A.~Kirillov,
Reidel: Dordrecht (1987);
R.~Casalbuoni, 
{\it On the quantization of systems with anti-commuting variables},
Il Nuovo Cimento {\bf 33A}, 115 (1976); 
{\it The classical mechanics for Bose-Fermi systems},
{\bf 33A}, 389 (1976);
A.~Barducci, R.~Casalbuoni and L.~Lusanna,
{\it Supersymmetry and the pseudoclassical relativistic electron},
{\bf 35A}, 377 (1976);
J.~Harnad and J.~P.~Par\'e, 
{\it Kaluza-Klein aproach to the motion of non-Abelian charged 
particles with spin}, 
Class. Quant. Grav. {\bf 8}, 1427 (1991).

\reference
V.~G.~Kats,
{\it Classification of simple Lie superalgebras},
Funct. Anal. Appl. {\bf 9}, 91 (1975);
L.~Corwin, Y.~Ne'eman and S.~Sternberg, 
{\it Graded Lie algebras in mathematics and physics (Bose-Fermi symmetry)},
Rev. Mod. Phys. {\bf 47}, p. 573 (1975);
B.~Kostant, {\it Graded manifolds, graded Lie theory, and prequantization},
in {\it Diff. Geom. Meths. in Math. Phys.} 
LNM {570}, p. 177, Springer: Berlin (1977);
R.~Giachetti, R.~Ragionieri, and R.~Ricci,
{\it Symplectic structures on graded manifolds},
Journ. Diff. Geom. {\bf 16}, 247 (1981);
A. El Gradechi, {\it On the Supersymplectic Homogeneous Superspace
Underlying the $\osp(1/2)$ coherent states},
Montr\'eal Preprint CRM-1850 (1993).

\reference 
R.~Jackiw,
{\it Dynamical symmetry of the magnetic monopole},
 Ann. Phys. (N.Y.) {\bf 129}, 183 (1980).

\reference E.~D'Hoker and L.~Vinet,
{\it Supersymmetry of the Pauli equation in the presence of a magnetic monopole},
Phys. Lett. {\bf 137B}, 72 (1984);
{\it Superspace formulation of the dynamical symmetries of the Dirac 
magnetic monopole},
Lett. Math. Phys. {\bf 8}, 439 (1984); 
{\it Dynamical supersymmetry of the magnetic monopole and the 
$1/r^2$-potential}
Comm. Math. Phys. {\bf 97}, 391 (1985);
E.~D'Hoker and L.~Vinet,
{\it Spectrum (super)symmetries of particles in a Coulomb potential},
Nucl. Phys. {\bf B260}, 79 (1985);
E. D'Hoker, V. A. Kosteleck\'y and L. Vinet, in
{\it Dynamical groups and spectrum generating algebras},
A. Bohm, Y. Ne'eman and A. O. Barut (eds), Vol. 1, p. 339;
Singapore: World Scientific (1988).

\reference
R.~Jackiw,
Ann.~Phys. (N.Y.) {\bf 201}, 83 (1990).

\reference
M.~De Crombrugghe and V.~Rittenberg, 
{\it Supersymmetric Quantum Mechanics},
Annals of Physics (N.Y.)
{\bf 151}, 99 (1983).

\reference
C.~J.~Park, Nucl. Phys. {\bf B376}, 99 (1992);
J.-G.~Demers, Mod. Phys. Lett. {\bf 8}, 827 (1993);
C.~Duval and P.~A.~Horvathy,
{\it Exotic supersymmetry of the magnetic vortex},
Tours Preprint 60/93 (unpublished).

\reference
A.~Pais and V.~Rittenberg, 
{\it Semisimple graded Lie algebras},
J. Math. Phys. {\bf 16}, 2062 (1975);
W.~Nahm, M.~Scheunert, 
{\it On the sructure of simple pseudo Lie algebras and their invariant 
bilinear forms}, J. Math. Phys. {\bf 17}, 868 (1976);
M.~Scheunert, W.~Nahm, V.~Rittenberg, 
{\it Irreducible representation of the $\osp(2,1)$ and ${\rm spl}(2,1)$ graded
Lie algebras}, J. Math. Phys. {\bf 18}, 155 (1977).

\reference
G.~Marmo, G.~Morandi, A.~Simoni and E.~C.~G.~Sudarshan, 
{\it Quasi-invariance and central extensions},
Phys. Rev. {\bf 37}, 2196 (1988).

\reference
R.~Puzalowski, {\it Galilean supersymmetry},
Acta Phys. Austriaca {\bf 50}, 45 (1978);
J.~A.~de Azc\'arraga and D.~Ginestar,
{\it Nonrelativistic limit of supersymmetric theories},
Journ. Math. Phys. {\bf 32}, 3500 (1991).

\bye